\documentclass[conference]{IEEEtran}

\usepackage{algorithm}
\usepackage{algorithmic}
\usepackage{amsmath}
\usepackage{color}
\usepackage{listings}
\usepackage{graphicx}
\usepackage{subfigure}

\definecolor{blue}{rgb}{0.098,0.357,0.675}
\definecolor{green}{rgb}{0.5,0.75,0.0}

\begin{document}

\title{Asynchronous Execution of the Fast Multipole Method Using Charm++}

\author{\IEEEauthorblockN{Mustafa AbdulJabbar, Rio Yokota, David Keyes}
\IEEEauthorblockA{King Abdullah University of Science \& Technology \\
 Thuwal, Saudi Arabia \\
\{mustafa.abduljabbar, rio.yokota, david.keyes\}@kaust.edu.sa}
}

\maketitle

\begin{abstract}
Fast multipole methods (FMM) on distributed memory have traditionally used a bulk-synchronous model of communicating the local essential tree (LET) and overlapping it with computation of the local data. This could be perceived as an extreme case of data aggregation, where the whole LET is communicated at once. \texttt{Charm++} allows a much finer control over the granularity of communication, and has a asynchronous execution model that fits well with the structure of our FMM code. Unlike previous work on asynchronous fast N-body methods such as \texttt{ChaNGa} and \texttt{PEPC}, the present work performs a direct comparison against the traditional bulk-synchronous approach and the asynchronous approach using \texttt{Charm++}. Furthermore, the serial performance of our FMM code is over an order of magnitude better than these previous codes, so it is much more challenging to hide the overhead of \texttt{Charm++}.
\end{abstract}

\section{Introduction}
When applying data-driven execution models to parallel hierarchical N-body methods, it is important first to understand the significance of the dynamic load-balancing and data prefetching mechanisms that have existed in them for over two decades. Parallel N-body methods start by partitioning the particles in a way that maximizes data locality while balancing the workload among the partitions. This is done by using the workload from the previous time step as weights when splitting a space filling curve that connects all particles \cite{Warren1993}. Parallel N-body methods also have a mechanism for prefetching the data on remote processes by communicating all necessary parts of the remote trees upfront. The resulting tree is a subset of the entire global tree, which is called the local essential tree (LET) \cite{Warren1993}. Any data-driven execution model that provides features such as dynamic load-balancing and data prefetching/caching must augment these existing tailored mechanisms rather than compete with them.

\begin{figure}[t]
\centering
\subfigure[Partitioning phase]{
\includegraphics[width=0.45\textwidth]{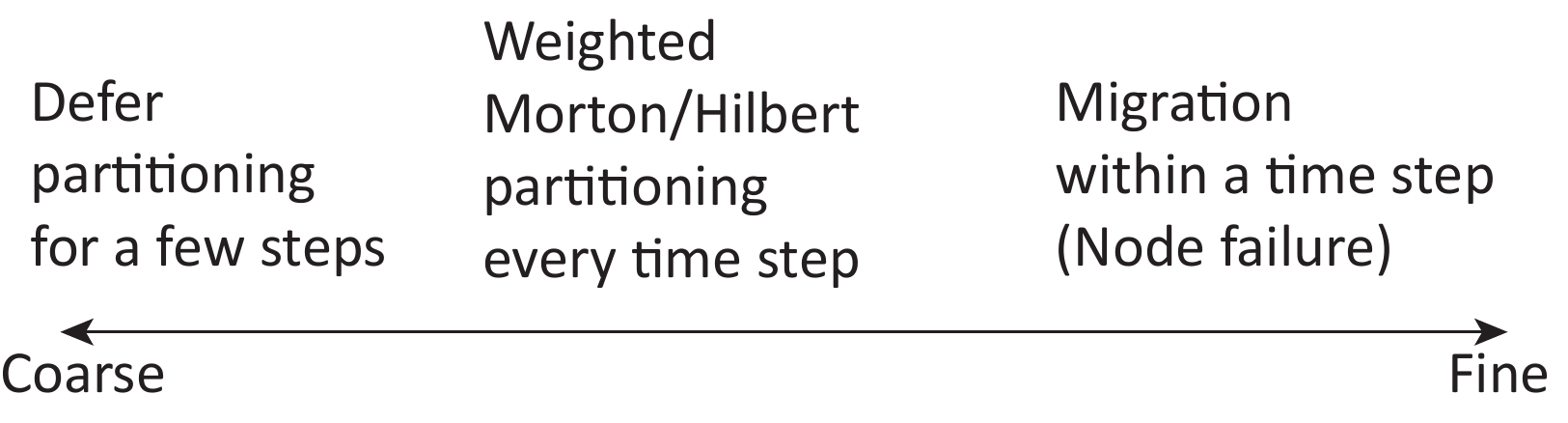}\label{fig:granularity_partition}}\\
\subfigure[LET communication phase]{
\includegraphics[width=0.45\textwidth]{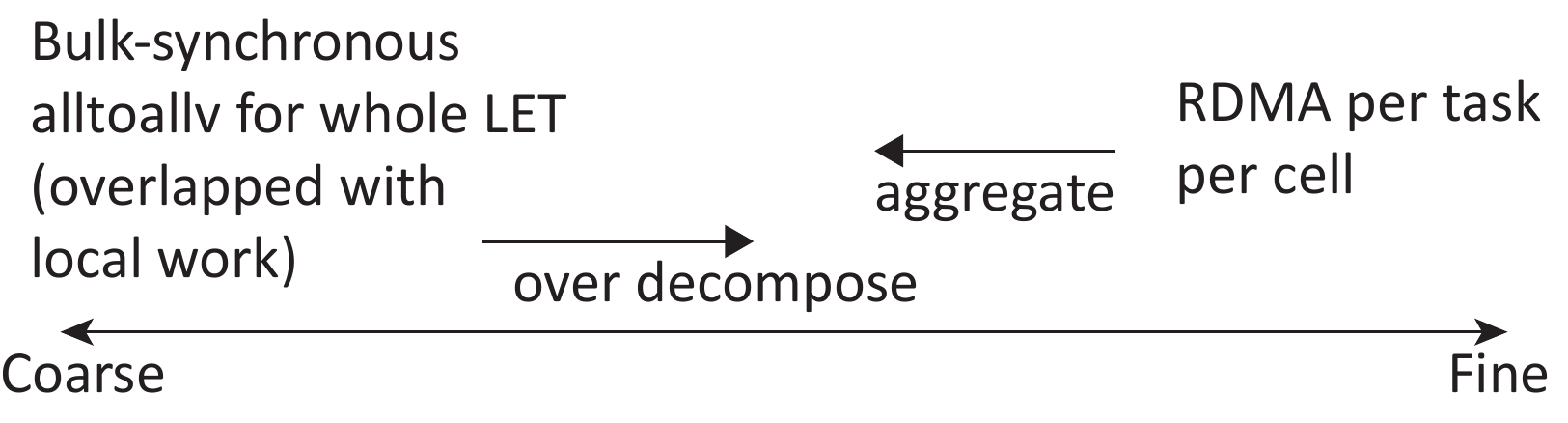}\label{fig:granularity_let}}
\caption{FMM has two major communication phases -- the partitioning of particles, and the communication of the local essential tree (LET). The former performs dynamic load-balancing and the latter can be thought of as a prefetching or data caching mechanism. Data-flow execution models add value not by providing these features, but by adding flexibility to the granularity at which these phases can be executed asynchronously. \ref{fig:granularity_let} shows the different granularity at which the partitioning phase can take place, while \ref{fig:granularity_let} shows the different granularity at which the LET communication can take place. The bulk-synchronous model can be viewed as an extreme case of communication aggregation.}
\label{fig:granularity}
\end{figure}

One area where the existing load-balancing and prefetching scheme can be improved is the granularity at which they are performed. Figure~\ref{fig:granularity_partition} shows the spectrum of granularity for the partitioning phase. Currently, the partitioning phase is constrained to the granularity of a single time step. One could coarsen the granularity by delaying the update of the partition for a few time steps, thereby adding more room for asynchronous execution. It is also possible that a repartitioning could take place within a time step in case of a node failure. Adding such flexibility to the partitioning granularity is a partial requirement for making the algorithm fault tolerant.

Figure~\ref{fig:granularity_let} shows the spectrum of granularity for the LET communication (prefetching) phase. Conventional parallel N-body methods use a bulk-synchronous \texttt{MPI\_alltoallv} to communicate the whole LET at once, and overlap this communication with the local tree traversal to hide latency. One could over decompose the LET down to a per cell request, and then aggregate the communication to the optimal granularity. The bulk-synchronous communication model can be thought of as an extreme case of aggregation, while something like an RDMA per task per cell would be at the other end of the granularity spectrum.

There have already been a few attempts to use data-driven execution models with parallel hierarchical N-body methods. Jetley \textit{et al.} use the \texttt{Charm++} execution model for their cosmological N-body code \texttt{ChaNGa} \cite{Jetley2008}. They compare several different cosmological datasets on several different architectures, and show significant improvement in the scalability over another cosmological N-body code \texttt{PKDGRAV}. They show that a na\"{i}ve load-balancing scheme based on work-stealing increases the amount of communication three-fold. \texttt{ChaNGa} has also been extended to run on GPUs \cite{Jetley2010}. The tree construction and tree traversal are done on the CPU and only the force calculation is performed on the GPU. They report 3.82 Tflops (single precision) on 896 CPU cores + 256 S1070 GPUs, which is less than 2\% of the theoretical peak. They are able to calculate approximately 10 million particles per second on 448 CPU cores + 128 GPUs. However, state-of-the-art parallel N-body codes such as \texttt{pfalcON} and \texttt{ExaFMM} can calculate 10 million particles per second on a single CPU socket \cite{Lange2014}.

When assessing the usefulness of new data-driven runtime systems, it is problematic to use a code with orders of magnitude slower serial performance. As mentioned earlier, data-driven execution models add value not by providing load-balancing or data-caching features to parallel N-body methods, but rather by adding flexibility to the granularity at which these mechanisms can be executed. However, slow serial performance of the code will skew the discussion on the optimal granularity. For example, techniques on the finer end of the spectrum in Figure~\ref{fig:granularity} will seem acceptable if the serial performance was slow enough, while in reality the communication latency could actually be too large for codes like \texttt{pfalcON} and \texttt{ExaFMM}. The same can be said to the case of Dekate \textit{et al.} \cite{Dekate2012}, where they use the \texttt{ParalleX} execution model for the Barnes-Hut treecode \cite{Barnes1986} and report a performance of 100K particles per second on a single CPU socket. This is exactly 100 times slower than the state-of-the-art N-body codes, which can do 10 million particle per second.

\begin{figure}[t]
\centering
\subfigure[Hierarchical interaction using FMM]{
\includegraphics[width=0.25\textwidth]{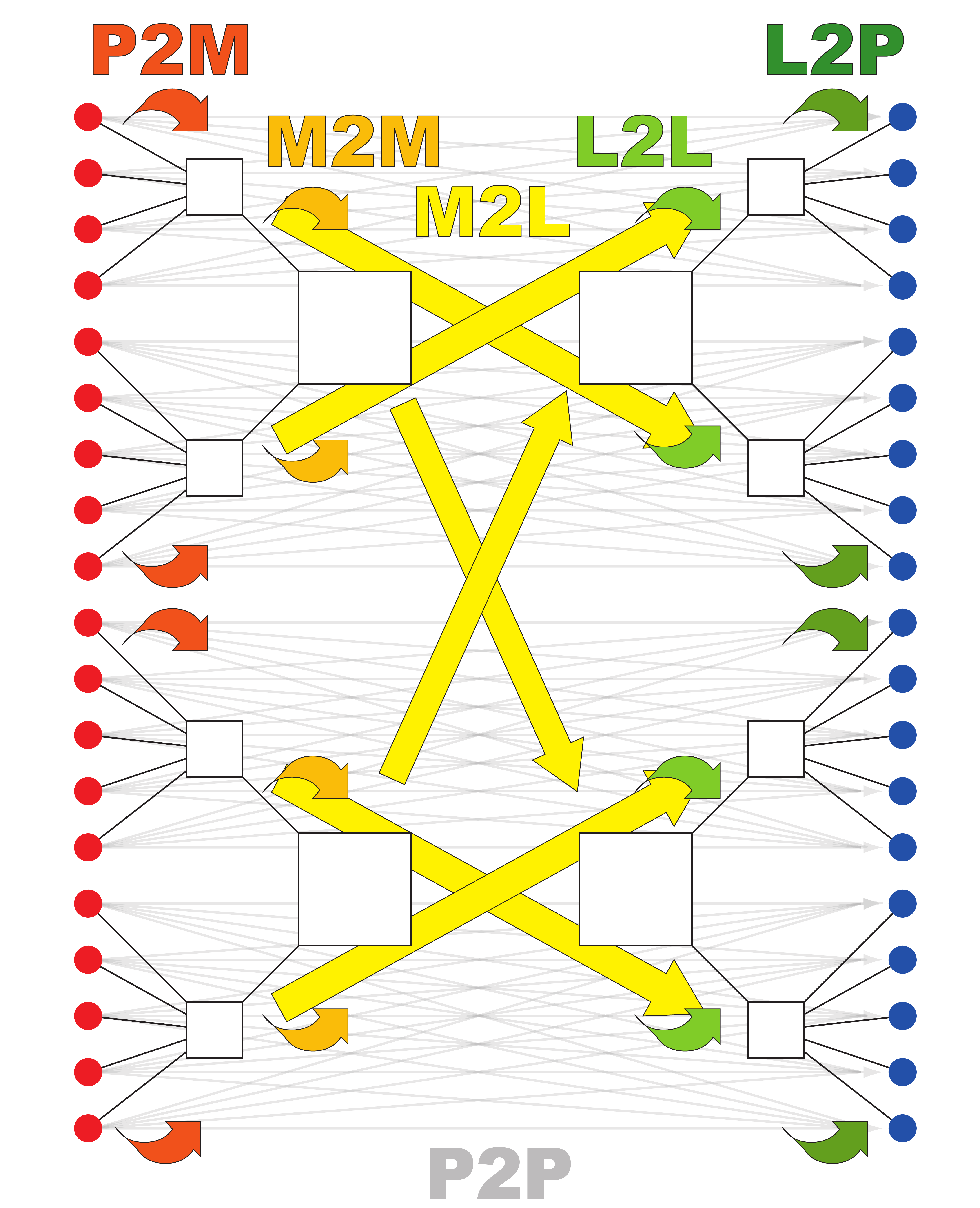}\label{fig:interaction}}\\
\subfigure[Flow of FMM calculation]{
\includegraphics[width=0.5\textwidth]{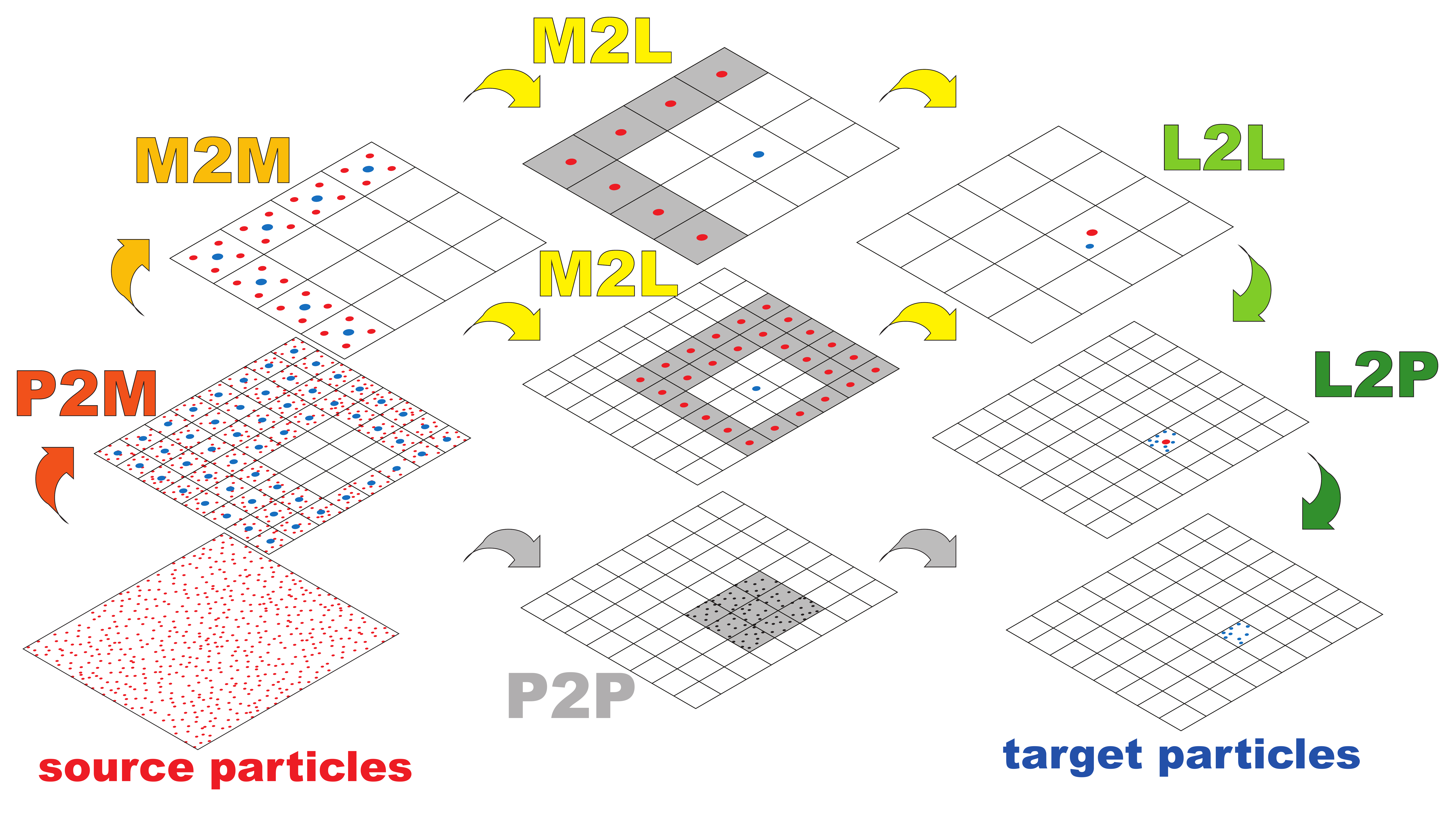}\label{fig:flow}}
\caption{Illustration of the flow of FMM calculation, and the interaction between source and target particles.}
\label{fig:fmm}
\end{figure}

There are a few other reports on the use of parallel N-body methods with data-driven execution models such as \texttt{StarPU} \cite{Agullo2012,Bordage2012} and \texttt{OmpSs}\cite{Pericas2012}, but these only consider shared memory architectures. Although there are qualitative similarities between inter-socket and inter-node data management, it is the quantitative difference that matters when discussing the granularity issues as mentioned before. The scope of the current work is on distributed memory data-driven execution models.

Previous work with good serial performance have focused on optimizing the bulk-synchronous all-to-all communication itself rather than data-driven execution models. With these optimizations Lashuk \textit{et al.} were able to calculate 90 billion particles in approximately 300 seconds on 200K cores of Jaguar and achieved 0.7 Pflops \cite{Lashuk2012}. Similarly, Yokota \textit{et al.} calculated 64 billion particles in approximately 100 seconds on 4000 GPUs of TSUBAME2.0 and achieved 1.0 Pflops \cite{Yokota2013}. The base of comparison for the data-driven execution models should be such highly optimized codes.

The present work performs a direct comparison between a highly scalable bulk-synchronous N-body code, \texttt{ExaFMM}, with and without \texttt{Charm++}. Unlike studies where the comparison is made against a completely different code, the present work compares the same code with and without the data-driven execution model.

\section{ExaFMM with Charm++}
\subsection{Data-flow of FMM}
In order to understand the amount of potential asynchronicity in FMM, one must understand the data-dependencies between the different phases of the FMM. FMM has six different mathematical operations that it performs, each with different dependencies and workloads. Figure~\ref{fig:fmm} shows the data-dependency of FMM in two separate schematics. The picture on the top shows a birds-eye view of the interaction between the red source particles and the blue target particles. Each connection or arrow shown in this figure represents a data-dependency. The interactions are grouped hierarchically so that far particles interact more sparsely. This is quite different from FFT, where even remote data points require equal amount of communication between them.

The bottom picture in Figure~\ref{fig:fmm} shows a geometrical partitioning of a two-dimensional domain and the corresponding data-dependency between the location of the cells. The P2M (particle-to-multipole) kernel takes the information of the particles (coordinates and charges) and calculates multipole expansion coefficients from this information (shown in red). For details of the mathematical formulation of FMM see \cite{Cheng1999, Cheng2006}. The M2M (multipole-to-multipole) kernel takes the information of the multipoles from its child cells and aggregates this information into a new multipole expansion at the center of a larger cell (shown in orange). The M2L (multipole-to-local) kernel takes all the information of the multipole expansions in the tree and translates them to local expansions (shown in yellow). There is a special rule for M2L kernels that it can only interact with cells that are sufficiently far compared to its cell size. The bigger the cell the further the other cell must be. The gray zones in Figure~\ref{fig:fmm} show the region where the M2L kernel is valid. The union of the gray zones on the three levels of M2L, M2L, and P2P shown in the center column add up to the entire domain. In a much deeper tree in 3-D this would be like peeling layers of M2L interaction lists until it reaches a ball of P2P neighbor lists, but they add up to the whole domain. The L2L (local-to-local) kernel takes the information given by the M2L kernel and cascades it down the tree (shown in light-green), until it reaches the bottom at which point the L2P (local-to-particle) kernel is called to translate that information to each particle (shown in dark-green).

The data-flow of FMM is analogous to a mail delivery system, where the information is aggregated from local post office to a larger hub, the delivery between remote locations is done with cargo aircraft, and then distributed back to the local post offices before delivery to the individual. This is a very efficient delivery system for computational problems that require information to travel from everywhere to every other place. Mathematically speaking, elliptical equations belong to this class of problems where some form of information must travel from one end of the domain to the other for the system to achieve a state of equilibrium. This could either be done by successive iteration with local halo communication over stencils, or it can be done more directly and asynchronously by packing/compressing everything and sending it over the network at once.

An important fact that can be overlooked by just looking at Figure~\ref{fig:fmm} is that the M2L and P2P phases are so much more expensive than the other phases. These two phases make up more than 90\% of the total runtime, so na\"{i}vely integrating the data-flow with the remaining 10\% will never result in a performance increase of more than 10\%. Furthermore, every M2L cell depends on hundreds of M2M cells. It is not a clear data-path where each node in the DAG has one arrow pointing to the next node. It is a DAG with hundreds of arrows pointing to a single M2L node. This is another reason why data-flow programming at this level of granularity is not favorable.

FMM has two major communication phases : the partitioning of particles (load-balancing), and the LET communication (prefetching). We describe in the next two subsections the details of each of these two phases and how \texttt{Charm++} is used to add asynchronicity and granular flexibility.

\subsection{Partitioning Phase}
\begin{figure}[t]
\centering
\subfigure[HOT (Morton)]{
\includegraphics[width=0.2\textwidth]{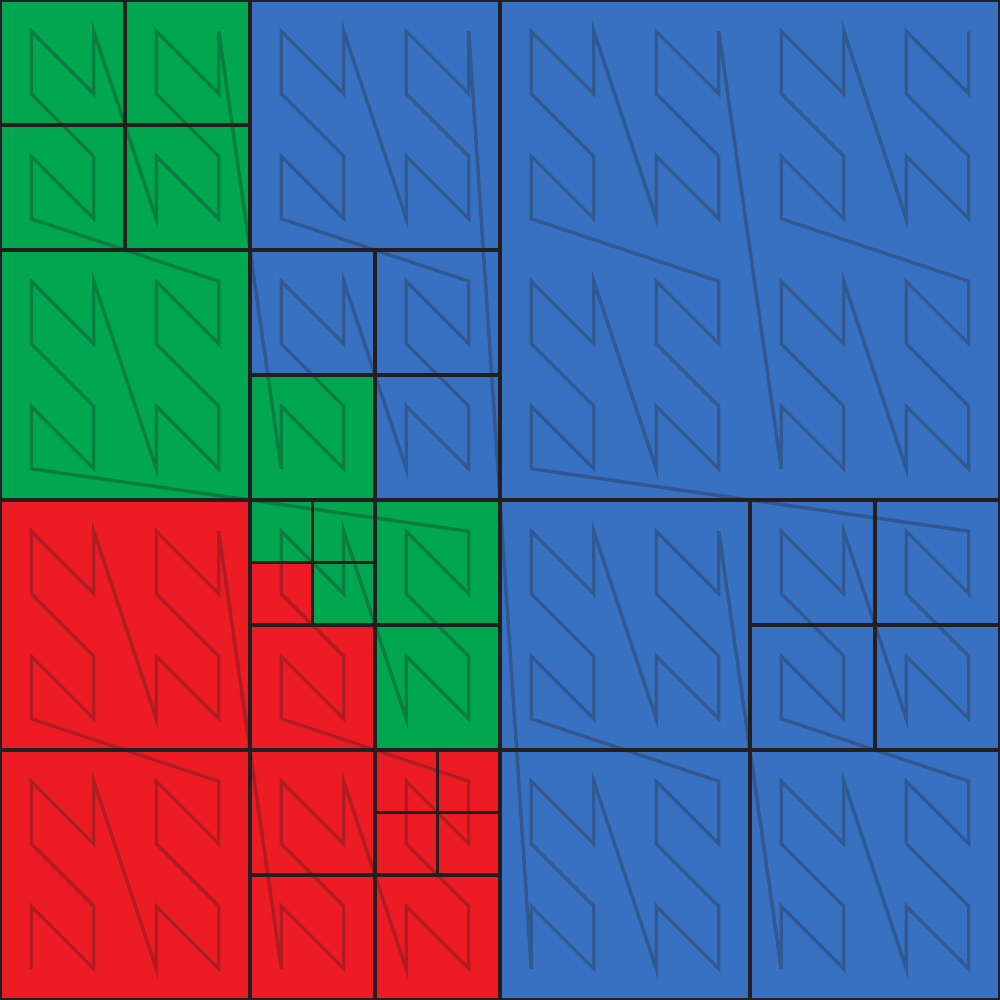}\label{fig:morton}}
\subfigure[HOT (Hilbert)]{
\includegraphics[width=0.2\textwidth]{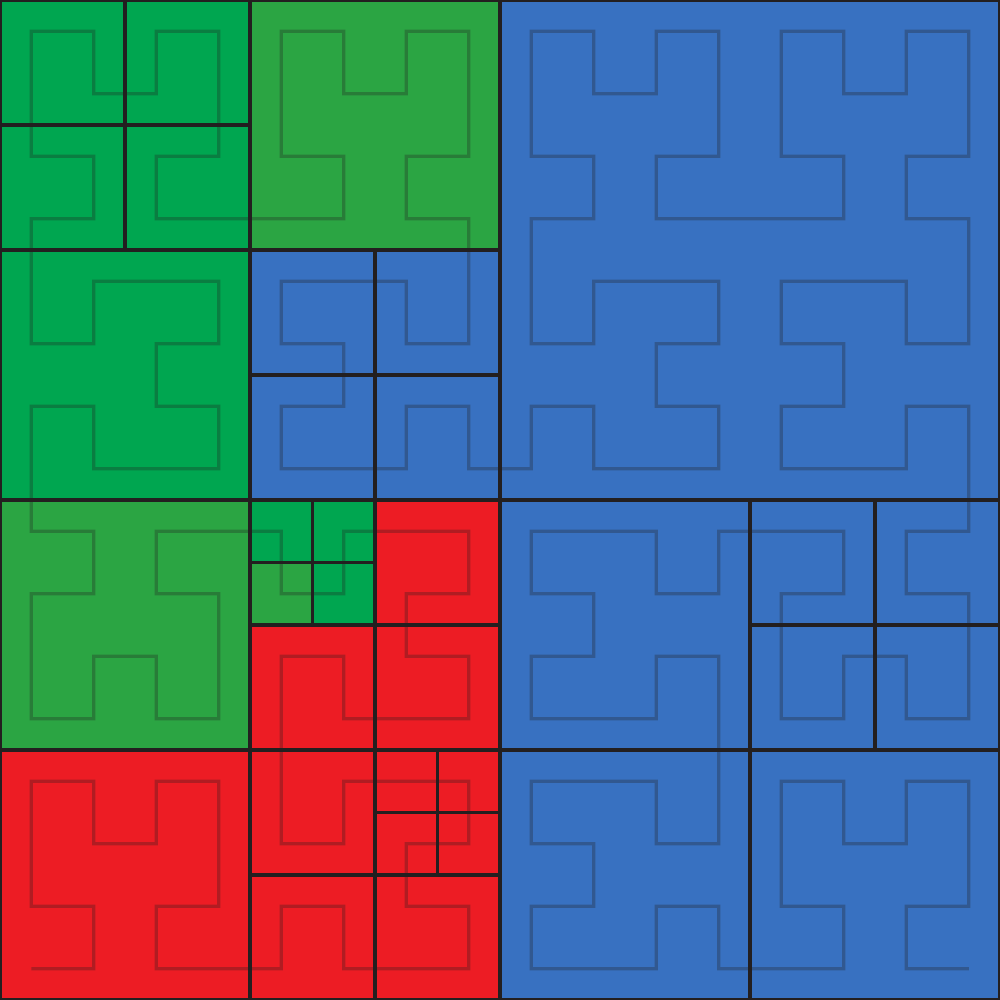}\label{fig:hilbert}}\\
\subfigure[ORB]{
\includegraphics[width=0.2\textwidth]{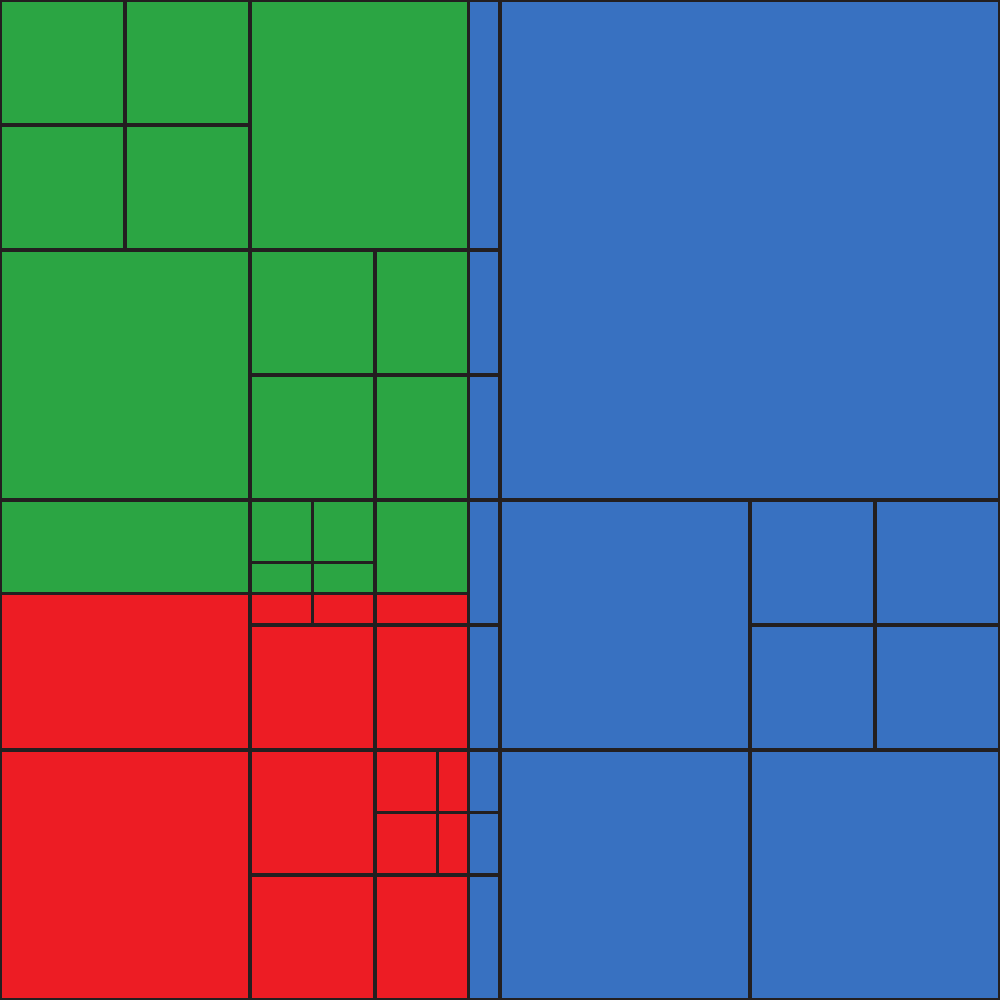}\label{fig:orb}}
\subfigure[Present method]{
\includegraphics[width=0.2\textwidth]{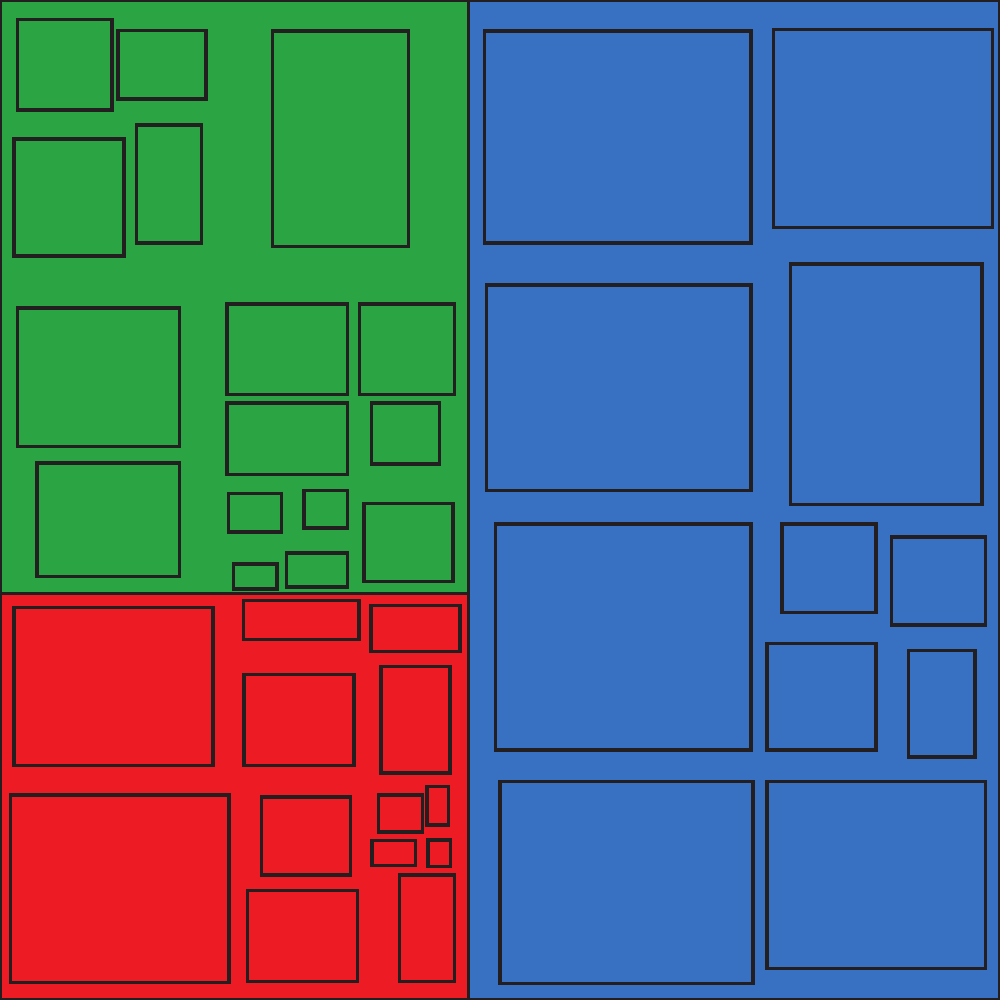}\label{fig:orb2}}\\
\caption{Schematic of different partitioning schemes. (a) shows the hashed octree with Morton keys. (b) shows the hashed octree with Hilbert keys. (c) shows the orthogonal recursive bisection with an underlying global tree. (d) is the present method using an orthogonal recursive bisection with independent local trees and tight bounding boxes.}
\label{fig:partition}
\end{figure}

Partitioning schemes for fast N-body methods can be categorized into orthogonal recursive bisections (ORB) \cite{Salmon1991} or hashed octrees (HOT) \cite{Warren1993}. The ORB forms a balanced binary tree by finding a geometric bisector that splits the number of particles equally at every bisection of the tree. The direction of the geometric bisector alternates orthogonally (x, y, z, x, ...) to form a cascade of rectangular subdomains that contain equal number of particles. For non-uniform distributions the aspect ratio of the subdomain could become large, which leads to suboptimal interaction list size and communication load. This problem can be solved by choosing the direction of the geometric bisector to always split in the longest dimension. The original method is limited to cases where the number of processes is a power of two, but the method can be extended to non-powers-of-two by using multi sections instead of bisections \cite{Makino2004}.

The HOT partitions the domain by splitting Morton/Hilbert ordered space filling curves into equal segments. Morton/Hilbert ordering maps the geometrical location of each particle to a single key. The value of the key depends on the depth of the tree at which the space filling curve is drawn. Three bits of the key are used to indicate which octant the particle belongs to at every level of the octree. Therefore, a 32-bit unsigned integer can represent a tree with 10 levels, and a 64-bit unsigned integer can represent a tree with 21 levels. Directly mapping this key to the memory address is inefficient for non-uniform distributions since most of the keys will not be used. Therefore, a hashing function is used to map the Morton/Hilbert key to the memory address of particles/cells.

\begin{figure}[t]
\centering
\includegraphics[width=0.45\textwidth]{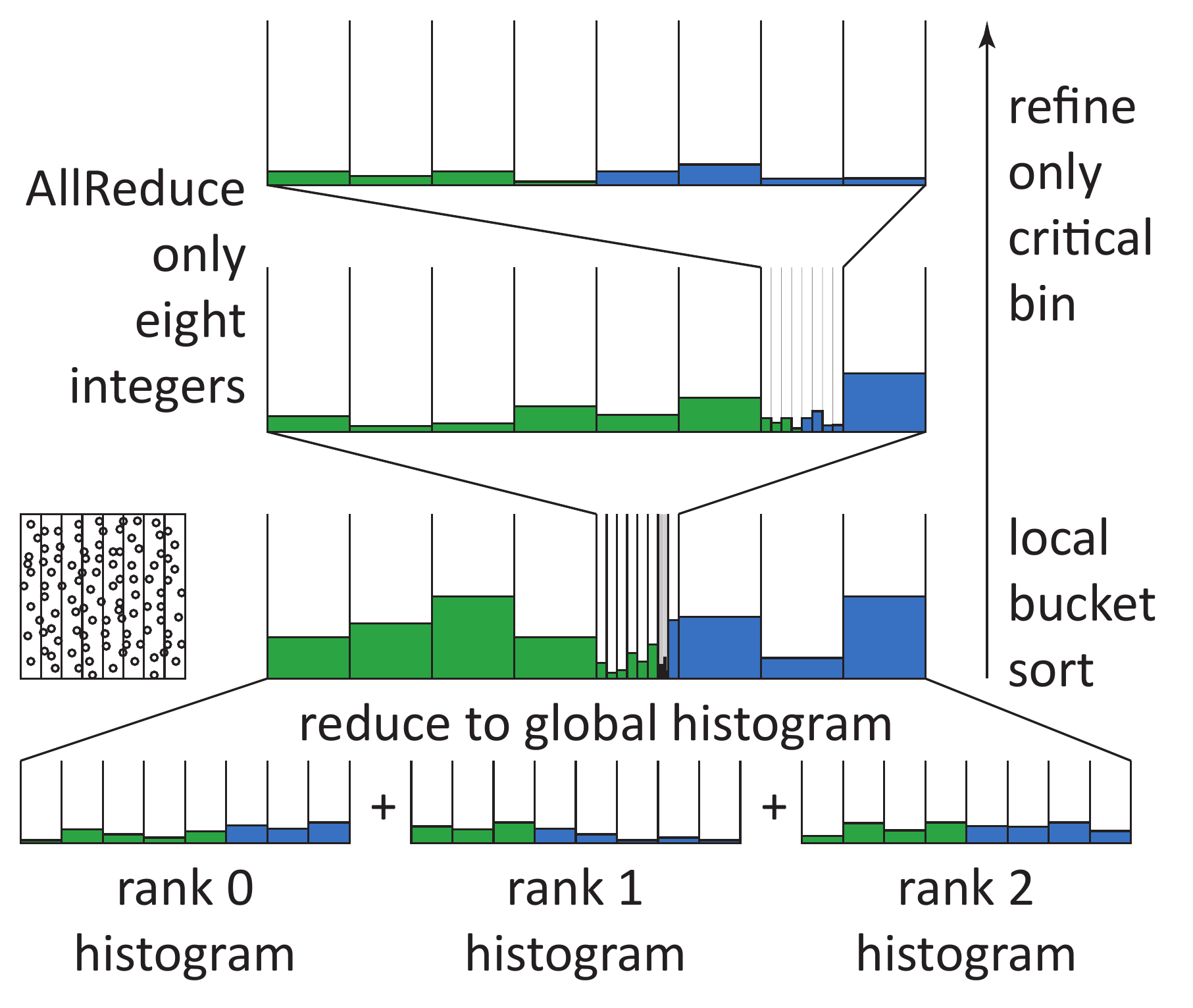}
\caption{Histogram-based partitioning scheme.}
\label{fig:split}
\end{figure}

Parallel sampling-based techniques have proven to be useful for both finding the bisectors in ORB \cite{Makino2004} and finding the splitting keys in HOT \cite{Solomonik2010}. Both ORB and HOT are constructing parallel tree structures, but in different ways. There is an analogy between parallel tree construction and parallel sorting. The idea behind ORB is analogous to merge sort, where a divide and conquer approach is taken. HOT is analogous to radix sort, where each bit of the key is examined at each step. Therefore, sampling-based techniques that are known to be effective for parallel sorting are also effective for parallel tree partitioning. The partitioning can be separated into two steps. The first step is to find the bisectors/key-splitters by using a sampling-based parallel sorting algorithm. An example of such sampling-based partitioning is shown in Figure~\ref{fig:split}. Sorting is only performed among the buckets (not within them) and this is done only locally. The only global information that is communicated is the histogram counts, which is only a few integers and can be done efficiently with an \texttt{MPI\_allreduce} operation. The bins can be iteratively refined to narrow the search for the splitter of the HOT key or ORB bisector. This will determine the destination process for each particle. The second step is to perform an all-to-all communication of the particles. Since the ORB bisector is one floating point number and the HOT key is one integer, it is much less data than sending around particle data at each step of the parallel sort.

In our current implementation we choose ORB over HOT for a few different reasons. One of the main reasons is that we were able to improve a major defect of ORB -- partition-cell alignment issue. Since geometrically closer points interact more densely with each other, it is crucial to keep the particles in the same cell on the same process in order to minimize communication. However, if a global Morton/Hilbert key is used to construct the local trees, the ORB may place a bisector in the middle of a cell as shown in Figure~\ref{fig:orb}. This results in an increase in the interaction list size. We avoid this problem by using local Morton/Hilbert keys that use the bounds of the local partition. This may at first seem to increase the interaction list near the partition boundaries since two misaligned tree structures are formed. However, when one considers the fact that the present method squeezes the bounding box of each cell to tightly fit the particles as shown in Figure~\ref{fig:orb2}, it can be seen that the cells are not aligned at all in the first place. Furthermore, our flexible definition of the multipole acceptance criteria optimizes the interaction list length for a given accuracy regardless of the misalignment.

There is one more important ingredient for an efficient partitioning scheme -- weighting by the workload. Particles have varying interaction list sizes, so equally splitting the bisection/key results in suboptimal load-balance. Weighting the particles with the workload from the previous time step is a simple and effective load-balancing strategy. This technique was mentioned in the original HOT paper \cite{Warren1993} and little has been done to improve it to this day. It would be na\"{i}ve to propose a work-stealing mechanism for fast N-body methods without understanding the significance of this practical solution that has stood the test of time. Data-driven execution models should be able to augment this tailored feature rather than to reinvent it. Although this weighting scheme was originally proposed for HOT, it can obviously be used to determine weights for particles during the bisection in ORB.

\begin{figure*}[t]
\centering
\subfigure[Conventional HOT/ORB]{
\includegraphics[width=0.9\textwidth]{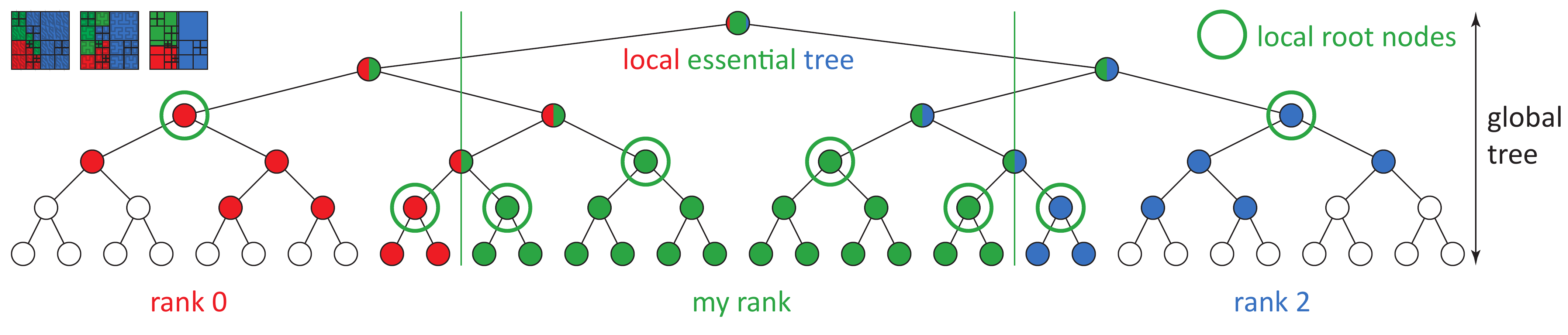}\label{fig:hot_tree}}\\
\subfigure[Present method]{
\includegraphics[width=0.9\textwidth]{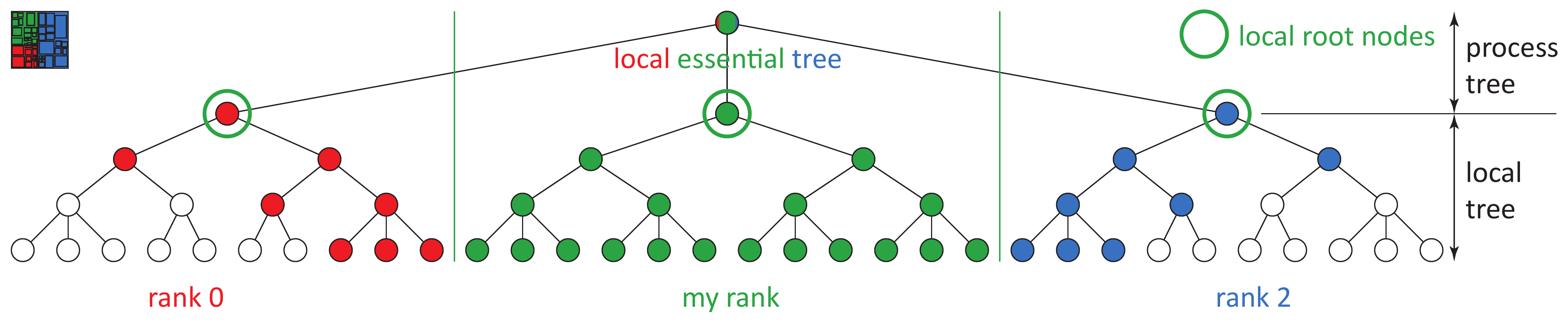}\label{fig:orb_tree}}
\caption{Schematic of the local essential tree (LET) for conventional HOT/ORB partitioning and present method.}
\label{fig:tree}
\end{figure*}

One limitation of the weighting scheme is that it only balances the workload and not the communication. There have been efforts to use graph partitioning tools with the workload as node-weights and communication as edge-weights, in order to create partitions that have an optimal balance of both the workload and communication \cite{Cruz2011}. This method has only been compared with Morton key splitting without weights, so the advantage over Morton key splitting with weights is unclear. In the present work we attempt to balance the workload and communication simultaneously by calculating the weight for the $i^{th}$ particle $w_i$ according to
\begin{equation}
w_i=l_i+\alpha*r_i
\label{eq:weight}
\end{equation}
where $l_i$ is the local interaction size, $r_i$ is the remote interaction list size, and $\alpha$ is a constant that is optimized over the time steps to minimize the total runtime. $l_i+r_i$ is the total interaction list size and represents the workload, while $r_i$ reflects the amount of communication. By adjusting the coefficient $\alpha$, one can amplify/damp the importance of communication balance. Making this an optimization problem to minimize the total runtime is what we prefer over minimizing the load-imbalance since the latter is not our final objective. Moreover, the variables $l_i$, $r_i$, and the total runtime are already measured in the present code so the information is available at no cost.

\texttt{Charm++} provides an opportunity to augment this bulk-synchronous approach to load-balancing by offering control over the granularity. We are interested in the data-flow programming model of \texttt{Charm++}, which allows us to asynchronously execute the FMM kernels while the communication for partitioning is happening. To our knowledge, there have not been any attempts to overlap computation with the communication in the partitioning phase. \texttt{Charm++} also provides task migration capabilities, but we decided not to use this feature for the current study. This is because we believe that it is much more efficient to ``strategically update" the partitions so that they are well balanced, than to try to ``steal work" after they are partitioned poorly.

\subsection{LET Communication Phase}
Once the particles are partitioned, the ones in the local domain are used to construct a local tree. We use a completely local construction of the octree using the local bounding box, instead of using a global Morton/Hilbert key that is derived from the global bounding box. This allows us to reuse all parts of the serial code and only add a few routines for the partitioning, grafting of trees, and communication. Therefore, any modification in the serial code is immediately reflected in the parallel code.

After the local tree structure is constructed, a post-order traversal is performed on the tree structure and P2M and M2M kernels are executed bottom up. The P2M kernel is executed only at the leaf cells. It loops over all particles in the leaf cell to form the multipole expansion at the center of the leaf cell. The M2M kernel is executed only for the non-leaf cells. It loops over all child cells and translates the multipole expansions from it's children's centers to its center.

Once the multipole expansions for all local cells have been determined, the multipole expansions are sent to the necessary processes in a sender-initiated fashion \cite{Dubinski1996}. This reduces the latency by communicating only once, rather than sending a request to remote processes and then receiving the data. Such sender-initiated communication schemes were common in cosmological N-body codes since they tend to use only monopoles, and in this case the integer to store the requests is as large as the data itself if they were to use a request-based scheme.

In the present method, the LET is formed from the information that is sent from the remote processes by simply grafting the root nodes of the remote trees as shown in Figure~\ref{fig:tree}. In conventional parallel FMM codes, a global octree is formed and partitioned using either HOT or ORB. Therefore, the tree structure was severed in many places as shown in Figure~\ref{fig:tree}, which caused the merging of the LET to become quite complicated. Typically, code for merging the LET would take a large portion of a parallel FMM code, and this made it difficult to implement new features such as periodic boundary conditions, mutual interaction, more efficient translation stencils, and dual tree traversals. \texttt{ExaFMM} is able to incorporate all these extended features and still maintain a fast pace of development because of this simplification in how the global tree structure is geometrically separated from the local tree structure.

While the remote information for the LET is being transferred, the local tree can be traversed. Conventional fast N-body methods overlap the entire LET communication with the entire local tree traversal. The LET communication becomes a bulk-synchronous \texttt{MPI\_alltoallv} type communication, where processes corresponding to geometrically far partitions send logarithmically less information, thus resulting in $\mathcal{O}(logP)$ communication complexity where $P$ is the number of processes. Nonetheless, in traditional fast N-body codes this part is performed in a bulk-synchronous manner.

\begin{figure}[t]
\begin{lstlisting}
entry void allToAllCells() {
  atomic {
    transportCellsToAll();
  }
  for (count=0; count<numChares; count++) {
    when transportCells(int cellCount, 
    Cell b[cellCount], int sender) atomic {
      processCells(cellCount, b, sender);
    }
  }	
  atomic {
    finishAllToAllCells();
  }
};
\end{lstlisting}
\caption{A generic bulk synchronous \texttt{Charm++}-mapped version of global communication }
\label{fig:alltoall}
\end{figure}

\texttt{ExaFMM}-\texttt{Charm++} replaces the global synchronization points, which consist of aggregation of LET data through all to all communication, with asynchronous sender-initiated entry functions that represent a coherent work entity called Chare. The local partitions are mapped to Chares that are accessible through entry functions. Such entry constructs propagate cell/body data across nodes to traverse the corresponding LET, rather than traverse all LETs in one shot. The advantages are reflected at both CPU and Network utilization levels. Global communication is typically synchronous in nature and are dependent on two factors: network bandwidth and initialization cost. It is clear that for large messages the first factor dominates whereas latency hiding could overcome this problem in rare cases; however, in general, such events act like global barriers. By supporting the traversal as an entry card to the LocalTree structure, the post-order traversal is triggered once the LET is received. The function in Figure~\ref{fig:alltoall} shows a direct mapping of the synchronized global communication to \texttt{Charm++}, that undergoes the Structured Control Flow abstraction. The ``when" construct controls the sequence at which messages are received, and the code inside it will be executed at the receiver side. It is clear from the abstraction that the next workload is triggered once all of messages are received. The function in Figure~\ref{fig:alltoall} is utilized by ExaFMM-Charm to replace the global blocking receive of cells with a remote asynchronous call that will process the message and proceed based on the rank of the sender.

Once the LET is formed the M2L and P2P kernels can be calculated using this information from the remote processes. The calculation of these two kernels takes a large portion of the execution time of FMM. The P2P kernel only requires information from its neighbors, while the M2L kernel requires information from an intermediate range. Besides these read-dependencies these two kernels do not have any level-wise dependency within the tree structure and can be processed in parallel on a per cell basis. In \texttt{ExaFMM} the M2L and P2P kernels are processes without forming an explicit interaction list by using the dual tree traversal \cite{Dehnen2002}. After the dual tree traversal is finished, a post-order traversal is performed and L2L and L2P kernels are executed to cascade the information down the tree to the particles.

\begin{table}[b]
\caption{FMM parameters}
\label{tab:parameters}
\begin{center}
\begin{tabular}{|c|c|c|c|c|c|c|}
\hline
Case & $N_{body}$ & $P$ & $\theta$ & $N_{crit}$ & $N_{spawn}$ & Distribution\\
\hline \hline
1 & $10^8$ & 10 & 0.4 & 256 & 1,000 & Cube\\
\hline
2 & $10^8$ & 10 & 0.4 & 512 & 1,000 & Sphere\\
\hline
3 & $10^8$ & 10 & 0.4 & 64 & 1,000 & Plummer\\
\hline
\end{tabular}
\end{center}
\end{table}

\section{Results}
Our tests were performed on the TACC Stampede system without using the coprocessors. Stampede has 6400 nodes, each with two Xeon E5-2680 processors with eight physical cores and 32GB of memory. We used the Intel compiler (\texttt{module intel/13.0.2.146, impi/4.1.0.030}) and used the Intel Thread Building Blocks library for the threading model. The \texttt{ExaFMM} code that was used for the current study is publicly available on bitbucket. \footnote{https://bitbucket.org/rioyokota/exafmm-dev}

\begin{figure}[t]
\centering
\subfigure[Cube]{
\includegraphics[width=0.2\textwidth]{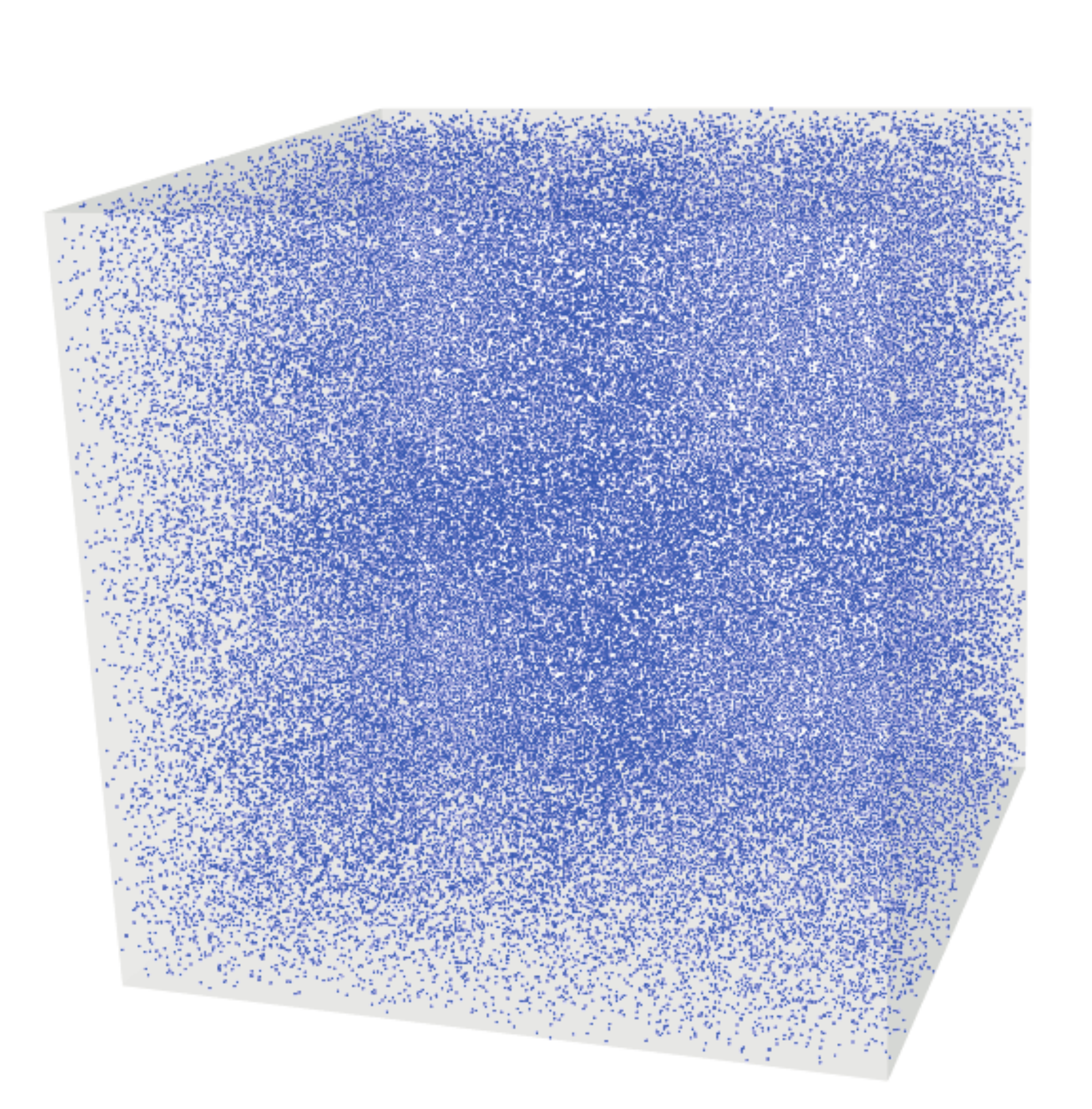}\label{fig:cube}}
\subfigure[Sphere]{
\includegraphics[width=0.2\textwidth]{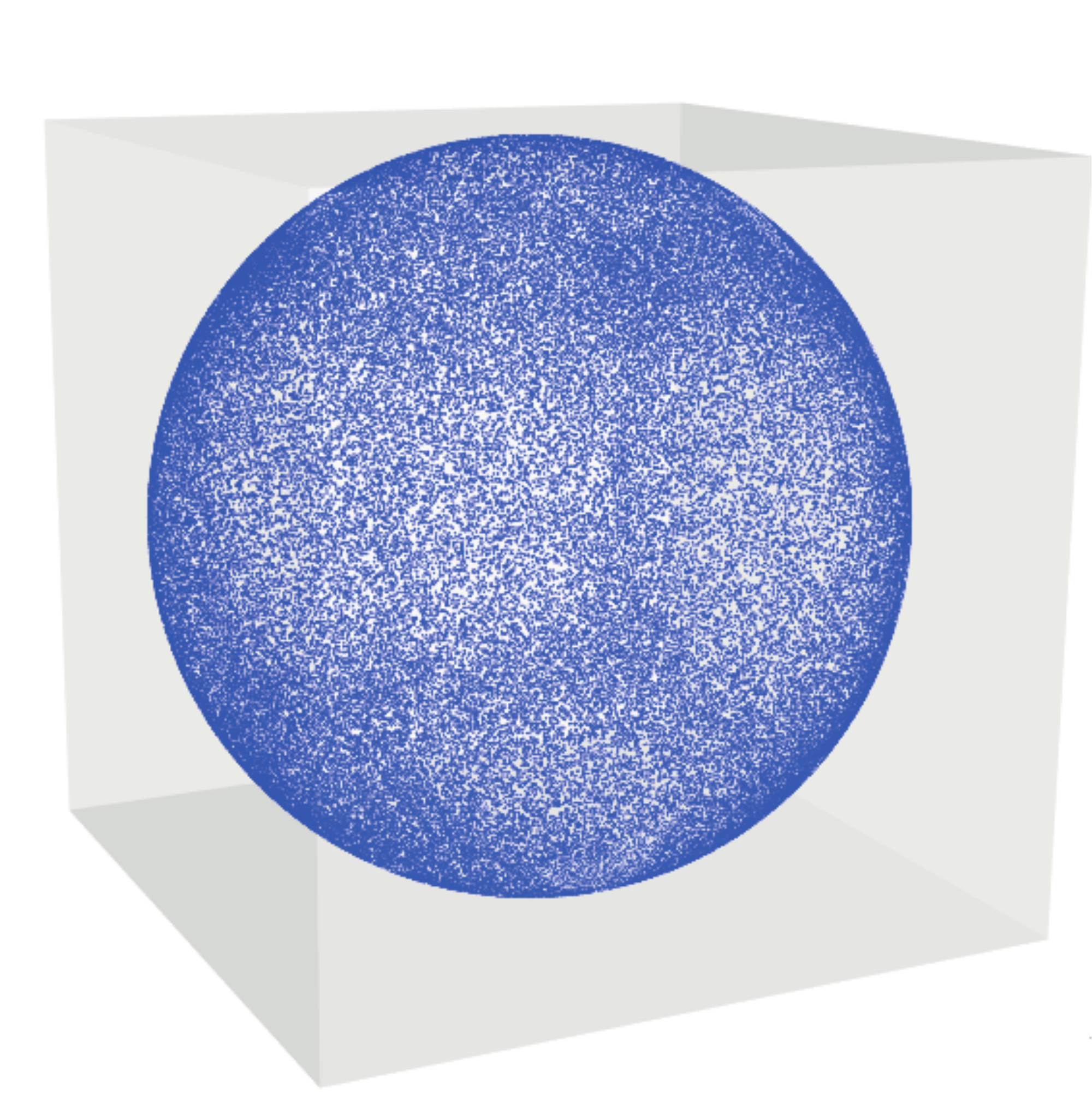}\label{fig:sphere}}
\subfigure[Plummer]{
\includegraphics[width=0.2\textwidth]{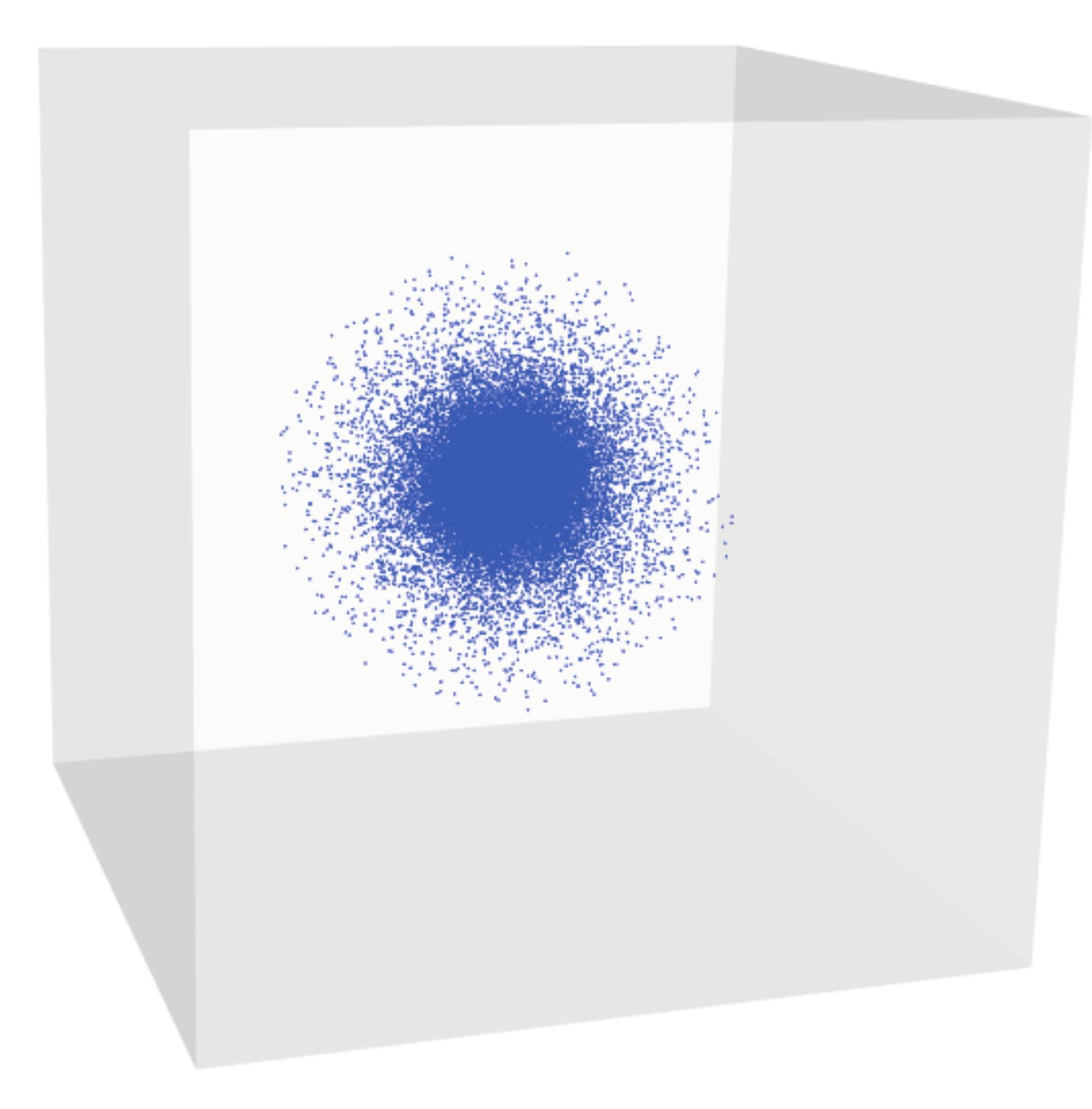}\label{fig:plummer}}
\caption{Different distributions of particles.}
\label{fig:distribution}
\end{figure}

\subsection{Strong Scalability}
We compare the scalability of \texttt{ExaFMM} with and without the use of \texttt{Charm++}. \texttt{ExaFMM} has many tunable parameters as shown in Table~\ref{tab:parameters}. $N_{body}$ is the number of bodies, $P$ is the order of multipole/local expansion, and $\theta$ is the multipole acceptance criteria. The FMM has a theoretical error bound of $\mathcal{O}(\theta^P)$, while the computational complexity varies between $\mathcal{O}(N\theta^{-2}P^3)$ and $\mathcal{O}(N\theta^{-2}P^6)$ depending on the type of basis \cite{Yokota2013a}. Unlike, previous treecodes that can only control $\theta$ or FMM codes that can only control $P$, \texttt{ExaFMM} can achieve the optimal speed by controlling both $P$ and $\theta$ simultaneously.

In Table~\ref{tab:parameters}, $N_{crit}$ represents the maximum number of particles per leaf cell, while $N_{spawn}$ is the minimum number of particles per spawned thread. Using a large $N_{crit}$ will create a shallower tree and decrease the number of M2L interactions, but will increase the number of particles per cell and therefore increase the number of P2P interactions. Using an optimal $N_{crit}$ value is essential to balance the workload between the M2L and P2P kernel, which are the two most expensive parts of the FMM. Increasing $N_{spawn}$ will allow less threads to be created and will decrease the overhead of the task spawning. Decreasing $N_{spawn}$ will cause more threads to be created and will make it easier to load-balance, but may increase the runtime due to the overhead caused by spawning many tasks. These values where carefully chosen to maximize the performance on Stampede.

\begin{figure}[t]
\centering
\includegraphics[width=0.45\textwidth]{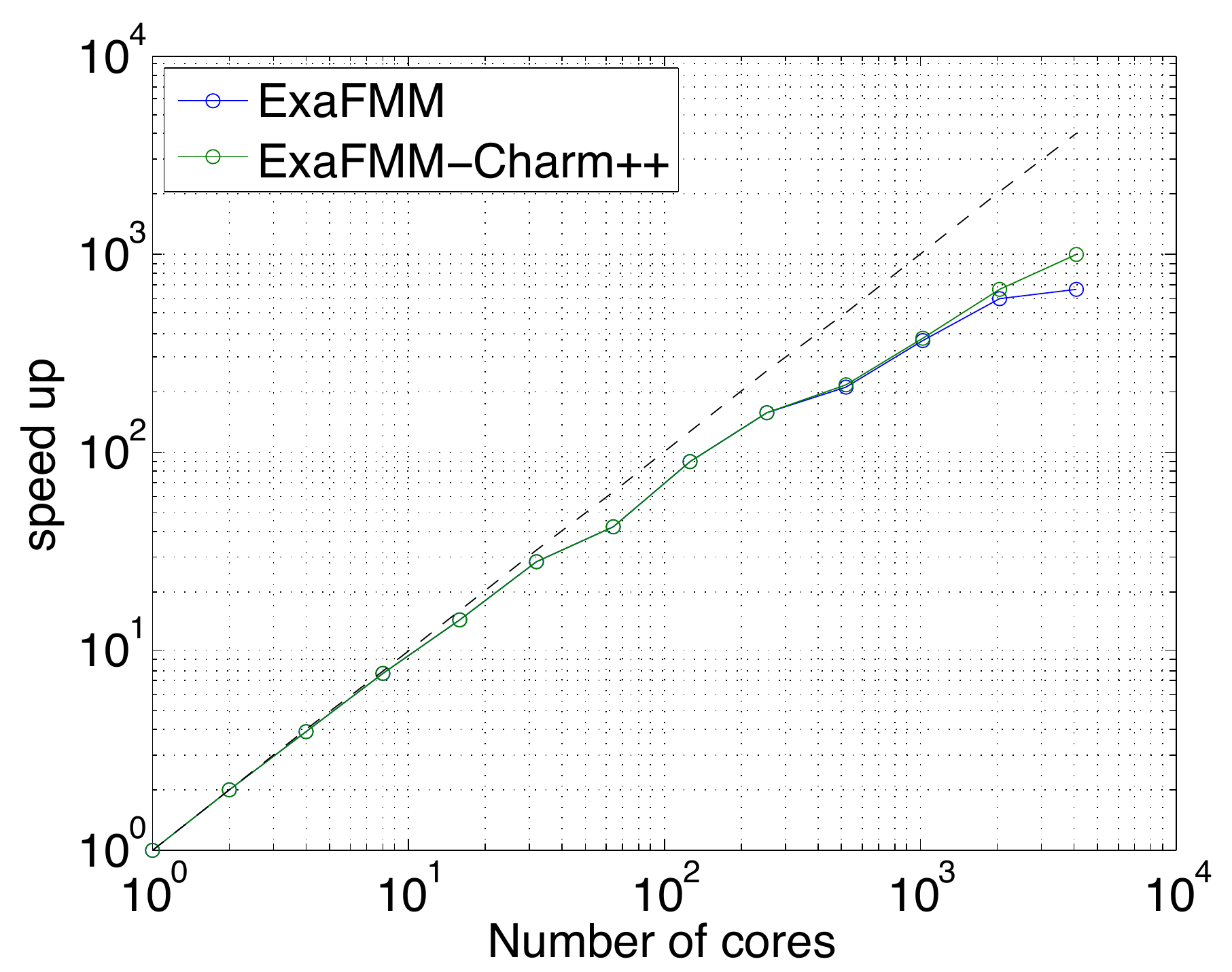}
\caption{Strong scaling for $N=10^8$ particles.}
\label{fig:scaling}
\end{figure}
The final entry in Table~\ref{tab:parameters} is the distribution of particles. We have selected three different types of distributions, which are representative of the actual distributions in scientific application codes. An illustration of the three distributions is shown in Figure~\ref{fig:distribution}. The ``Cube" distribution can be found in molecular dynamics simulations where water molecules are evenly distributed throughout a cubic domain. The ``Sphere" distribution has points only on the surface of the sphere. This is representative of boundary integral problems, where a surface mesh is used to discretize the problem. The ``Plummer" distribution is typical for cosmological N-body simulations, where the mass is distributed unevenly with very high concentration in certain areas.

We perform a strong scalability test of the FMM by keeping the number of particles to the value shown in Table~\ref{tab:parameters} and increasing the number of cores. All the values in Table~\ref{tab:parameters} are kept constant throughout the strong scalability tests. We first run up to 16 cores per node and then increase the node count once the number of cores per node is saturated. The total number of cores used in the largest run was 4,096.

The results of the strong scalability test using \texttt{ExaFMM} with and without \texttt{Charm++} are shown in Figure~\ref{fig:scaling}. The divergence from ideal scaling is mainly caused by the increase in the interaction list size when splitting the constant-sized tree into smaller and smaller segments. By looking back at Figure~\ref{fig:partition}, one can see that all partitioning schemes will suffer from this problem because it is difficult to maintain a small surface to volume ratio when partitioning a constant domain into thousands of subdomains. For any partitioning scheme, the shapes of the partitions tend to be neater at a macroscopic level, but the unevenness in the particle distribution at the microscopic scale tends to create oddly-shaped partitions as you go finer.

\begin{figure}[t]
\centering
\includegraphics[width=0.45\textwidth]{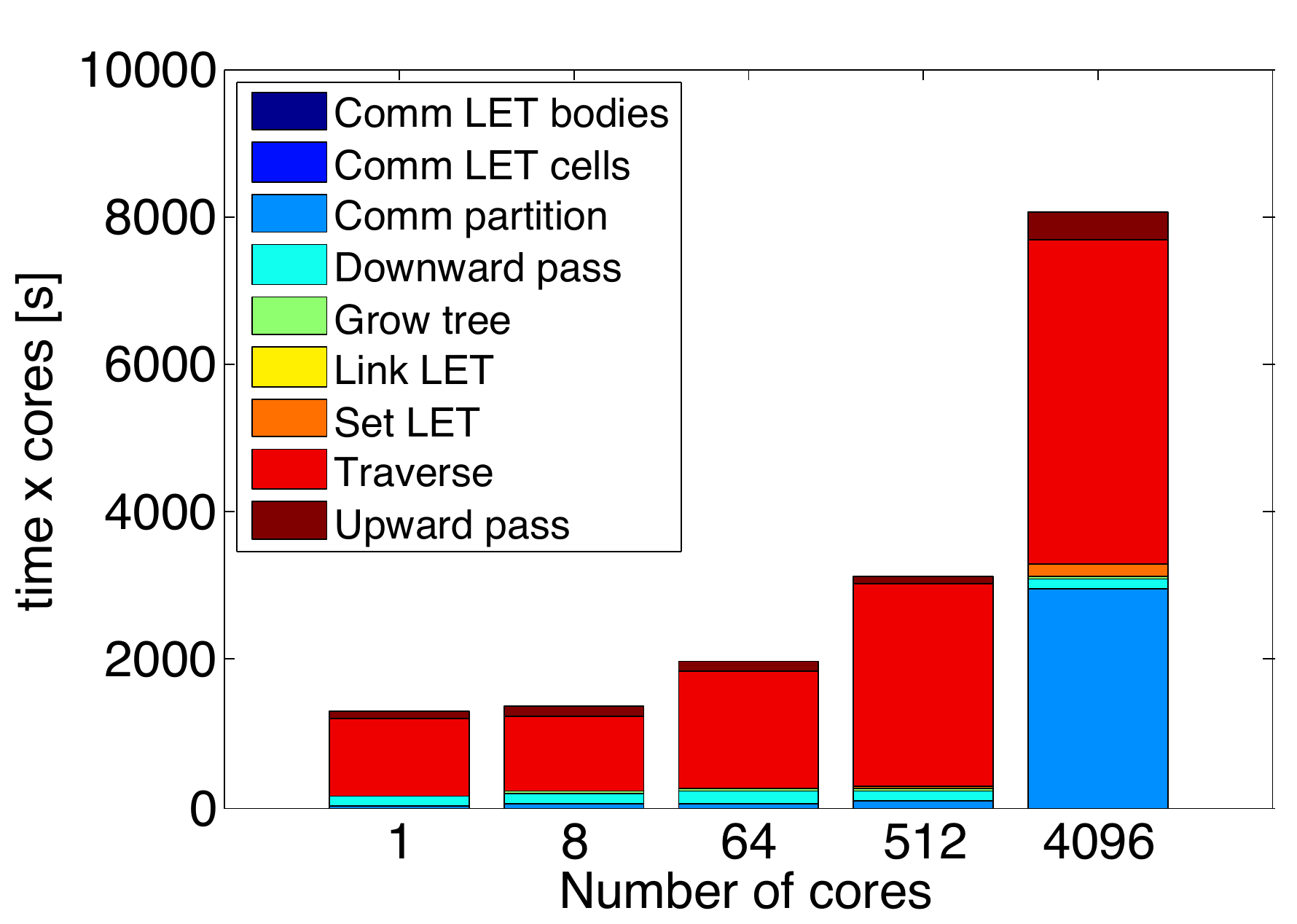}
\caption{Breakdown of strong scaling for $N=10^8$ particles.}
\label{fig:breakdown}
\end{figure}

We will take a closer look at the strong scalability runs by plotting the breakdown of the runtime in Figure~\ref{fig:breakdown}. The breakdown in Figure~\ref{fig:breakdown} corresponds to the plot for ``ExaFMM" in Figure~\ref{fig:scaling}. The main difference between the two plots is that Figure~\ref{fig:scaling} is showing the speedup, whereas Figure~\ref{fig:breakdown} is showing the runtime multiplied by the number of cores. This is done so that the bar plot for larger core counts is clearly visible. Therefore, in Figure~\ref{fig:breakdown} a constant bar height will mean perfect strong scalability. 

It can be seen that the ``Comm partition" phase is consuming a large time on 4,096 cores. This is the communication of the partitioning stage, which is very large for the initial step. Note that the two LET communication phases ``Comm LET bodies" and ``Comm LET cells" are completely overlapped with the ``Traverse" and cannot be seen in Figure~\ref{fig:breakdown}. The original \texttt{ExaFMM} code overlaps the LET communication with local tree traversal so adding \texttt{Charm++} does not improve the performance any further for this part. However, the communication for the initial partitioning phase is not overlapped with any computation in \texttt{ExaFMM} (or any other fast N-body code as far as the authors are aware), so the asynchronous execution model of \texttt{Charm++} provides some benefit for this part.

\subsection{Load Balancing}
\begin{figure}[t]
\centering
\includegraphics[width=0.45\textwidth]{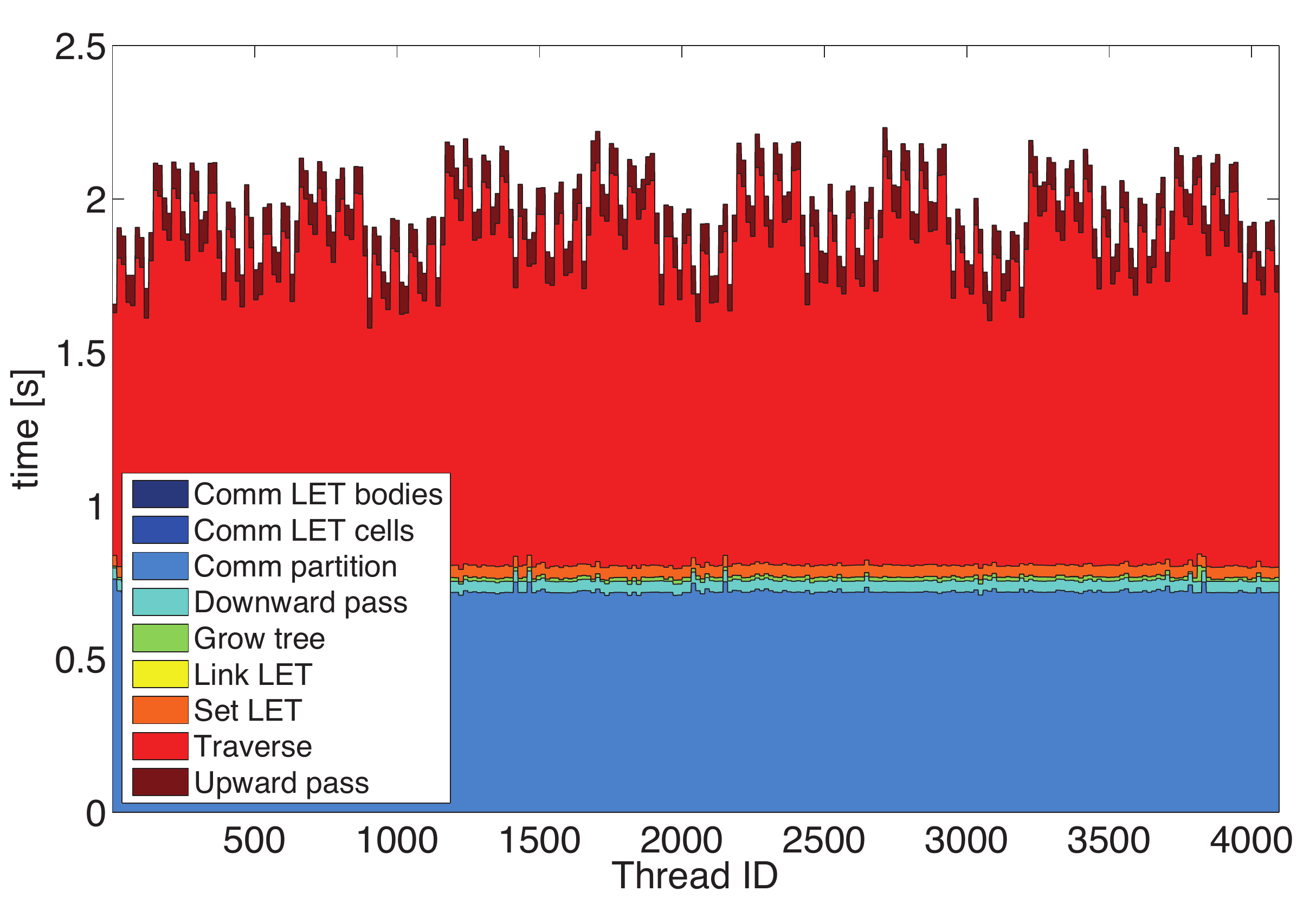}
\caption{Load-balance for the Plummer distribution.}
\label{fig:loadbalance}
\end{figure}

The increase in runtime of the ``Traverse" phase shown in Figure~\ref{fig:breakdown} is mostly attributed to the increase in the interaction list length as mentioned earlier, but it is also partially caused by load-imbalance. We see this in Figure~\ref{fig:loadbalance}, where the runtime across all cores is shown with the same legend as Figure~\ref{fig:breakdown} but this time without multiplying the runtime by the number of cores. As can be seen from Figure~\ref{fig:distribution} the Plummer distribution is highly non-uniform and is difficult to partition to thousand of subdomains. Furthermore, the main difficulty of partitioning N-body codes is that the work-load is not directly proportional to the partition size. For mesh-based methods, partitioning into equal size subdomains would result in somewhat equal workload. However, since each particle has a different interaction list size, partitioning the domain so that the number of particles are equal will result in suboptimal load-balance.

\begin{figure*}[t]
\centering
\subfigure[Cube]{
\includegraphics[width=0.32\textwidth]{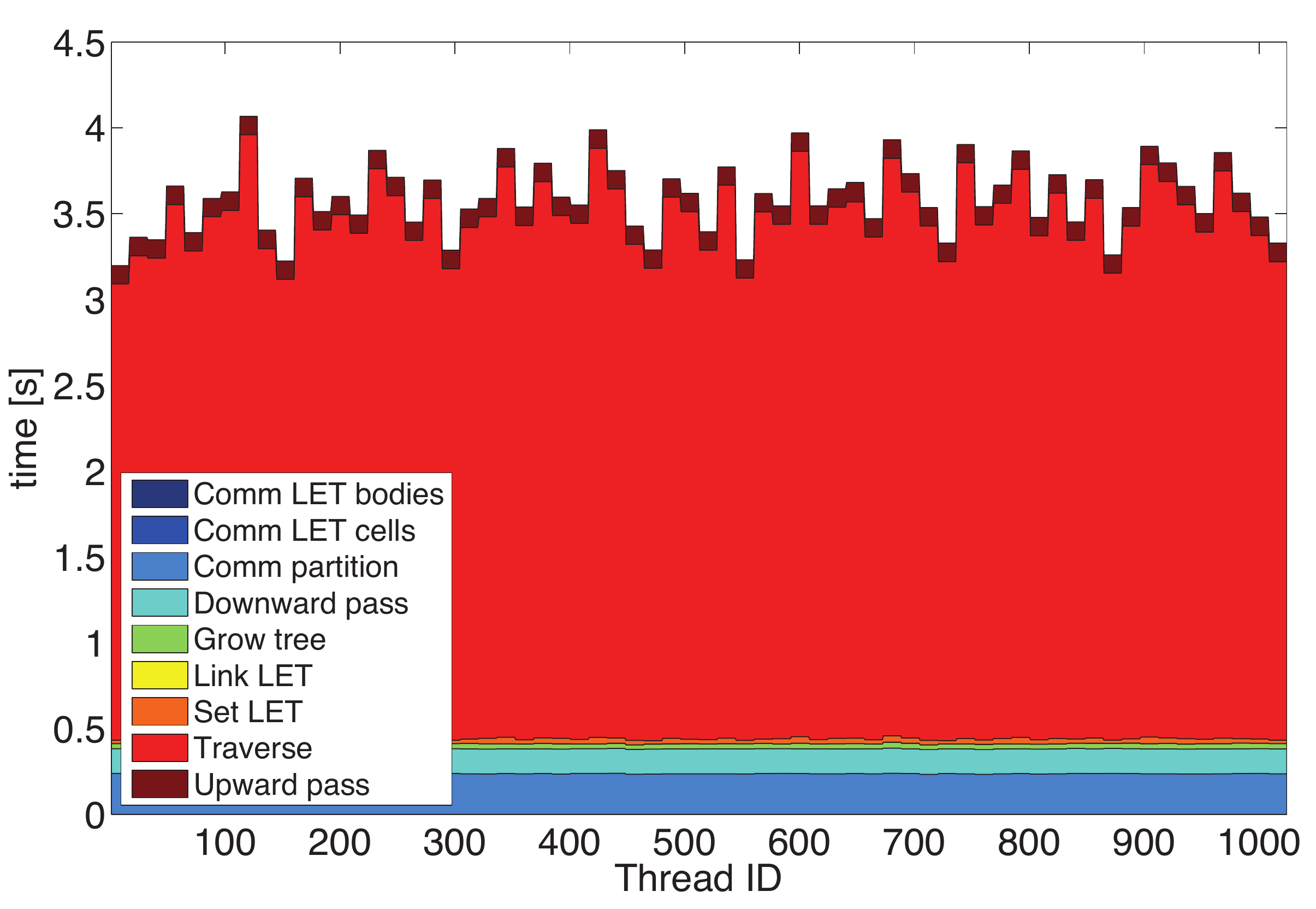}\label{fig:cube1}}
\subfigure[Sphere]{
\includegraphics[width=0.32\textwidth]{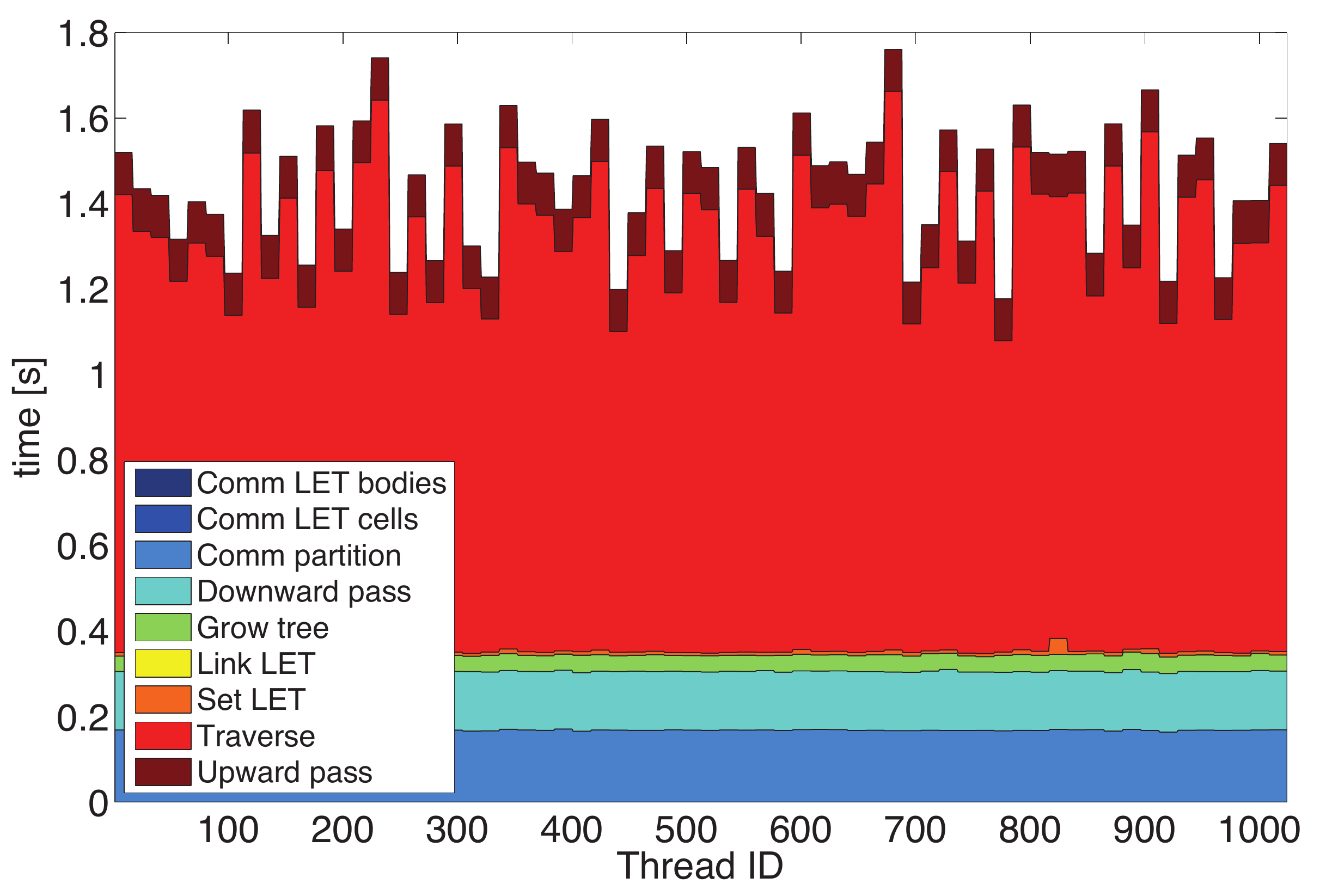}\label{fig:sphere1}}
\subfigure[Plummer]{
\includegraphics[width=0.32\textwidth]{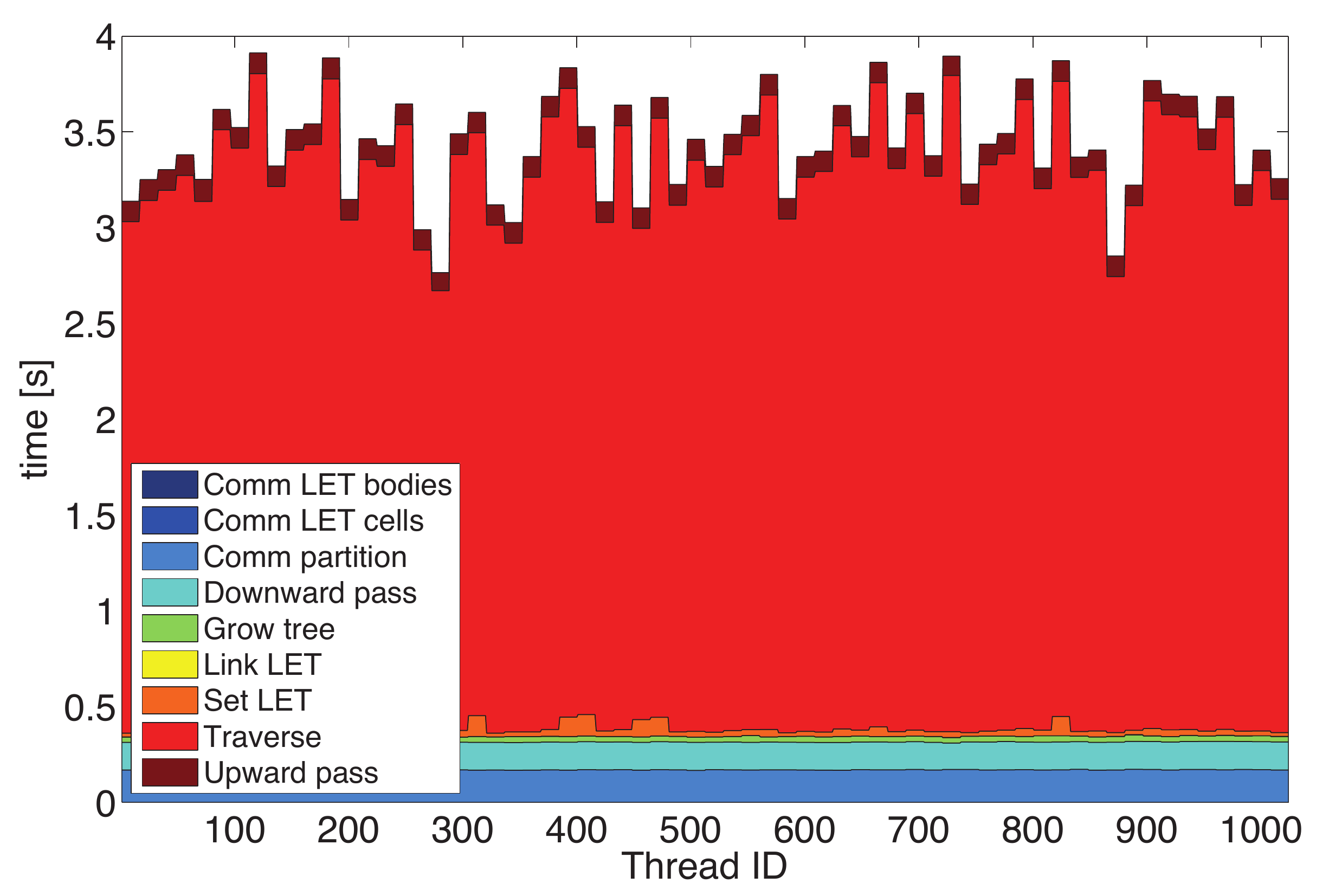}\label{fig:plummer1}}\\
\subfigure[Cube (new weighting scheme)]{
\includegraphics[width=0.32\textwidth]{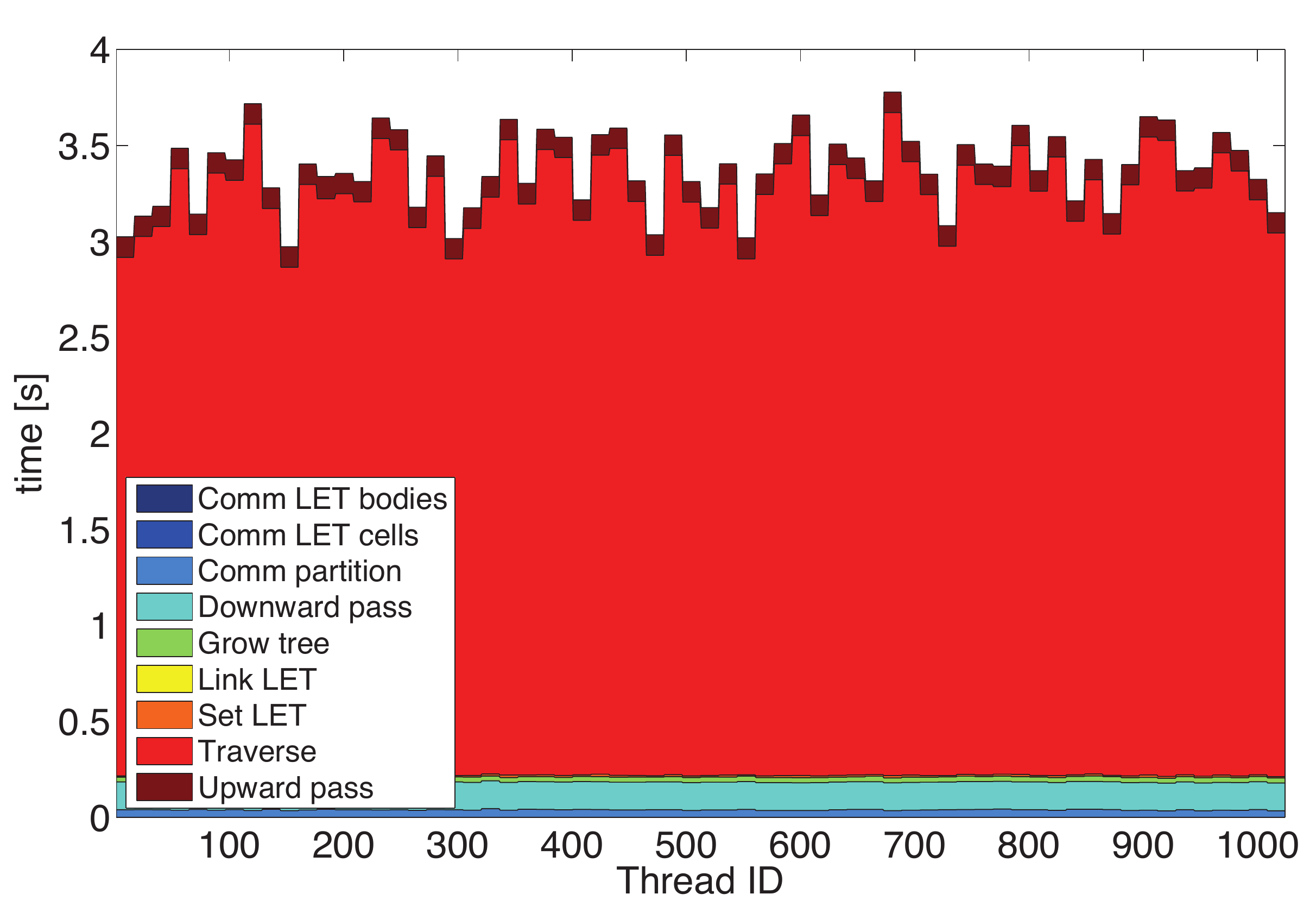}\label{fig:cube1}}
\subfigure[Sphere (new weighting scheme)]{
\includegraphics[width=0.32\textwidth]{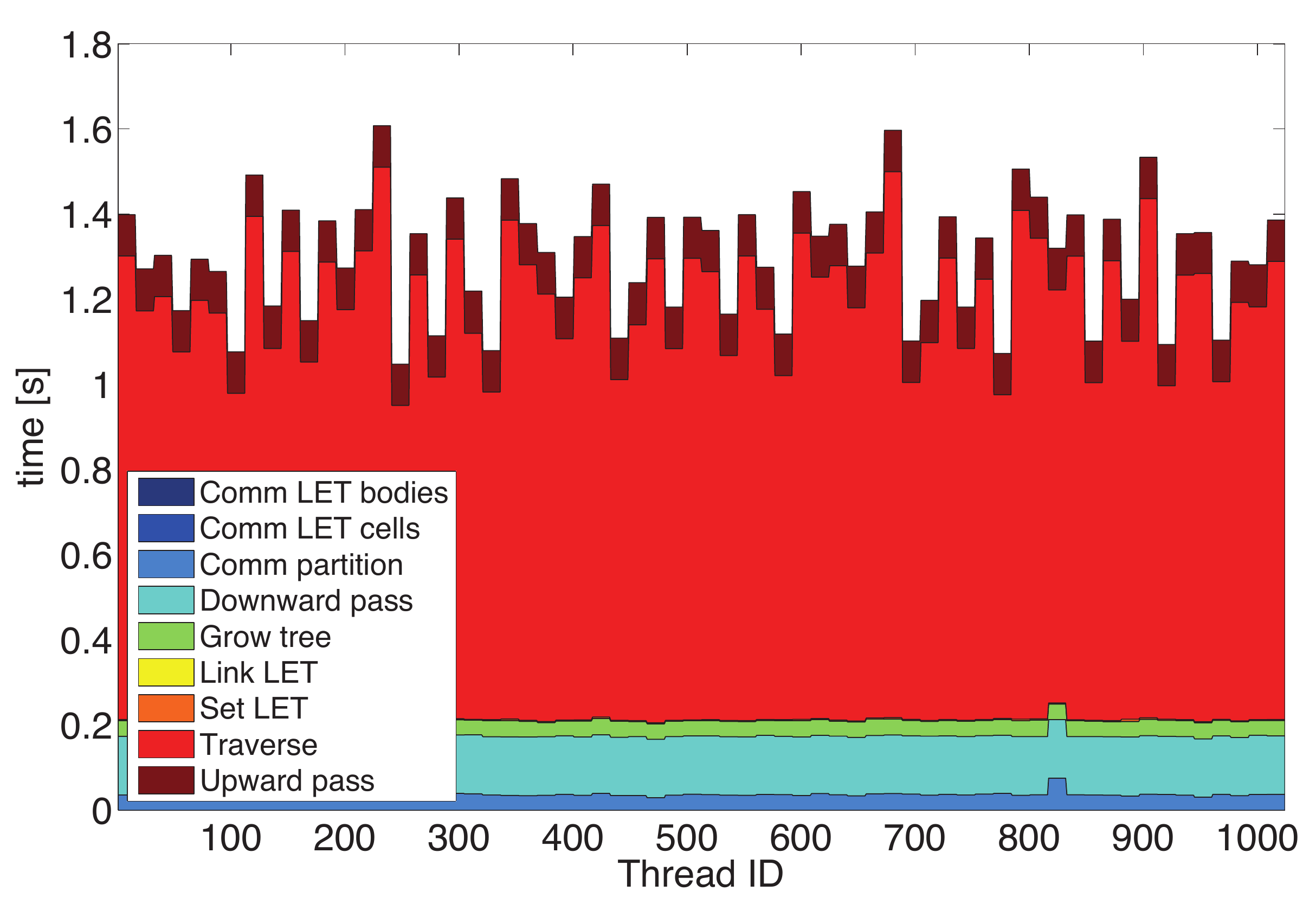}\label{fig:sphere1}}
\subfigure[Plummer (new weighting scheme)]{
\includegraphics[width=0.32\textwidth]{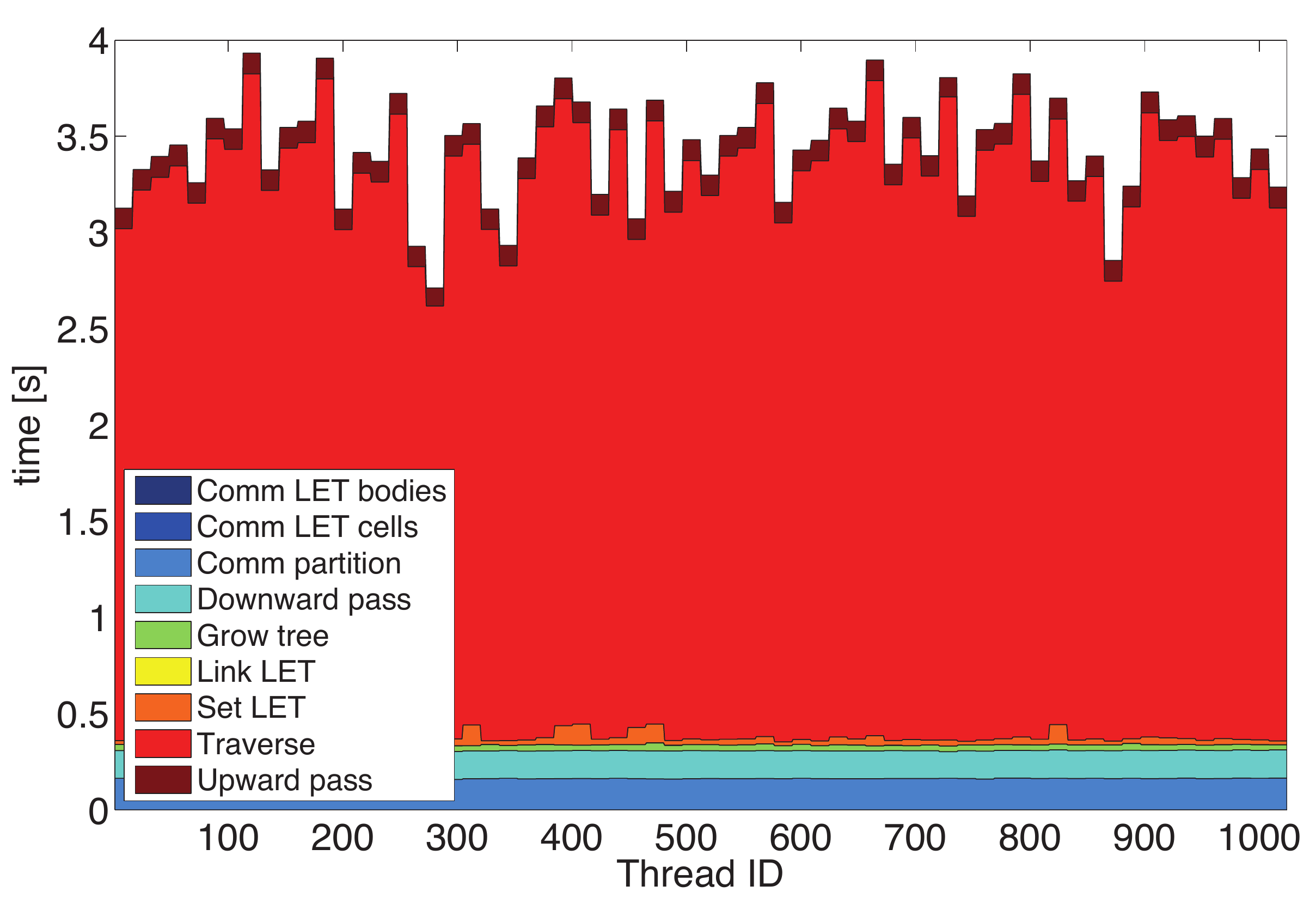}\label{fig:plummer1}}
\caption{Different distributions of particles. $N=10^8$ on $1,025$ (not $1,024$) cores}
\label{fig:weights}
\end{figure*}

It is difficult to assess the quality of our load-balancing scheme by just looking at a single case, especially if perfect balance is not a reasonable goal to aim for. Therefore, we will compare the results for the different distributions shown in Table~\ref{tab:parameters}. If the most benign distribution is showing the same amount of load-imbalance as the most difficult case. Then we should be able to conclude that there is little room for improvement. Figure~\ref{fig:weights} shows the distribution of the runtime on each core for six different cases. The top three cases are for the ``Cube", ``Sphere", and ``Plummer" distribution with the standard weighting scheme based on the interaction list length. The bottom three cases are for the same distributions but with our new weighting scheme that tries to optimize for both the work and communication load by using Eq.~\eqref{eq:weight}. The number of cores is set to 1,025 to create an environment where the number of processes is not a power of two. We used 16 threads per node except for the last node which used only 1 thread. Therefore, we had to use one extra node, which makes the partitioning of MPI ranks not so straightforward.

We see from Figure~\ref{fig:weights} that all distributions have somewhat similar load imbalance despite these difficult conditions. Therefore, we conclude that our partitioning scheme can handle difficult distributions and difficult number of processes to the same degree that it can handle the easy ones.

\section{Conclusions}
Distributed memory parallelization models for FMM have traditionally been bulk-synchronous, but dynamic load-balancing and data prefetching mechanisms have existed in them for over two decades. For load-balancing, the hashed-octree and orthogonal recursive bisection are both effective techniques for maximizing data locality while balancing the workload among the partitions by using the workload from the previous step as weights when partitioning. For data-prefetching, the local essential tree is formed by communicating all necessary parts of the remote tree upfront. These two techniques are usually applied at the granularity of the time step, but the data-flow of FMM allows a more flexible granularity for both load-balancing and data prefetching. We have investigated the possibility of using \texttt{Charm++} to have a finer control over the granularity of the communication and asynchronous execution.

Unlike previous work on asynchronous fast N-body methods such as \texttt{ChaNGa} and \texttt{PEPC}, the present work performs a direct comparison against the traditional bulk-synchronous approach and the asynchronous approach using \texttt{Charm++}. Furthermore, the serial performance of our FMM code is over an order of magnitude better than these previous codes, so it is much more challenging to hide the overhead of \texttt{Charm++}.

We also propose a novel partitioning scheme, which allows us to geometrically separate the local tree from the global tree. This was only possible because our FMM uses the dual tree traversal, which does not require a global key nor cubic cells. By taking advantage of this feature of the dual tree traversal, we were able to simplify the grafting of the local essential tree greatly. This simplification of our code made it possible to readily integrate with frameworks such as \texttt{Charm++} with relative ease.

In order to demonstrate the effectiveness of the current combination of state-of-the-art load-balancing, data-prefetching, and data-flow execution models, we performed a strong scalability test that spans over three orders of magnitude without offsetting the problem size. As expected, the communication for the initial partitioning phase became a bottleneck at 4096 cores, but we were able to improve this by using asynchronous execution model of \texttt{Charm++}. This allows us to achieve over a 1000 times speedup for an highly non-uniform Plummer distribution with $N=10^8$ particles.

We confirmed that our weighting scheme for the partitioning works evenly well for various particle distributions. Random distribution in a cube, points on a spherical shell, and the highly non-uniform Plummer distribution all had a similar load-imbalance for $N=10^8$ particles on 1025 (not 1024) cores. This also demonstrates that our partitioning scheme works equally well for non-powers of two. The \texttt{ExaFMM} code that was used for the current study is publicly available on bitbucket. \footnote{https://bitbucket.org/rioyokota/exafmm-dev}

\section*{Acknowledgments}
This publication was based on work supported in part by Award No KUK-C1-013-04, made by King Abdullah University of Science and Technology (KAUST). This work used the Extreme Science and Engineering Discovery Environment (XSEDE), which is supported by National Science Foundation grant number OCI-1053575.

\bibliographystyle{IEEEtran}

\end{document}